\pdfoutput=1

\documentclass[11pt,twoside,a4paper,cmspaper,final,collab]{cms-tdr}
\def\svnVersion{3e2cfe0-D}\def\svnDate{2020/08/24}\def\cmsCernNoTag{CERN-EP-2020-064}\def\cmsCernDate{\today}\def\cmsMessage{"Published in Physics Letters B as \href{http://dx.doi.org/10.1016/j.physletb.2020.135710}{\doi{10.1016/j.physletb.2020.135710}.}"}

\begin{document}\cmsNoteHeader{SMP-19-012}

\hyphenation{had-ron-i-za-tion}
\hyphenation{cal-or-i-me-ter}
\hyphenation{de-vices}
\newcommand{\mll}{\ensuremath{m_{\ell\ell}}\xspace}
\newcommand{\mlll}{\ensuremath{m_{\ell\ell\ell}}\xspace}
\newcommand{\jet}{\ensuremath{\mathrm{j}}}
\newcommand{\mjj}{\ensuremath{m_{\jet\jet}}\xspace}
\newcommand{\detajj}{\ensuremath{\abs{\Delta\eta_{\jet\jet}}}\xspace}
\newcommand{\dphijj}{\ensuremath{\Delta\phi_{\jet\jet}}\xspace}
\newcommand{\WW}{\ensuremath{\PW^\pm\PW^\pm}\xspace}
\newcommand{\WZ}{\ensuremath{\PW\cPZ}\xspace}
\newcommand{\tZq}{\ensuremath{\PQt\cPZ\Pq}\xspace}
\newcommand{\tVx}{\ensuremath{\PQt\PV\mathrm{x}}\xspace}
\newcommand{\ptmax}{\ensuremath{\pt^{\text{max}}}\xspace}
\newcommand{\zepmax}{\ensuremath{\max(z_{\ell}^{*})}\xspace}
\providecommand{\bigabs}[1]{\ensuremath{\Bigl|#1\Bigr|}\xspace}
\newcommand{\MocaPlus}{\textsc{MoCaNLO}+\textsc{Recola}\xspace}

\newlength\cmsTabSkip\setlength{\cmsTabSkip}{1ex}
\ifthenelse{\boolean{cms@external}}{\providecommand{\cmsTable}[1]{#1}}{\providecommand{\cmsTable}[1]{\resizebox{\textwidth}{!}{#1}}}
\ifthenelse{\boolean{cms@external}}{\providecommand{\cmsLeft}{upper\xspace}}{\providecommand{\cmsLeft}{left\xspace}}
\ifthenelse{\boolean{cms@external}}{\providecommand{\cmsRight}{lower\xspace}}{\providecommand{\cmsRight}{right\xspace}}

\cmsNoteHeader{SMP-19-012}

\title{Measurements of production cross sections of $\WZ$ and same-sign $\PW\PW$ boson pairs in association with two jets in proton-proton collisions at $\sqrt{s}=13\TeV$}

\author*[inst1]{CMS experiment}

\date{\today}

\abstract{
Measurements of production cross sections of $\WZ$ and same-sign $\PW\PW$ boson pairs in association with two jets in proton-proton collisions 
at $\sqrt{s}=13\TeV$ at the LHC are reported. The data sample corresponds to an 
integrated luminosity of 137\fbinv, collected with 
the CMS detector during 2016--2018. The measurements are performed in the leptonic decay modes 
$\PW^\pm\cPZ \to \ell^\pm\PGn\ell'^\pm\ell'^\mp$ and 
$\WW \to \ell^\pm\PGn\ell'^\pm\PGn$, where $\ell, \ell' = \Pe$, 
$\PGm$. Differential fiducial cross sections as functions of the invariant masses of the jet 
and charged lepton pairs, as well as of the leading-lepton transverse momentum, are 
measured for $\WW$ production and are consistent with the 
standard model predictions. The dependence of differential cross sections 
on the invariant mass of the jet pair is also measured for $\WZ$ production. 
An observation of electroweak production of $\WZ$ boson pairs is reported 
with an observed (expected) significance of 6.8 (5.3) standard deviations. 
Constraints are obtained on the structure of quartic vector boson interactions 
in the framework of effective field theory.}

\hypersetup{%
pdfauthor={CMS Collaboration},%
pdftitle={Measurements of production cross sections of WZ and same-sign WW boson pairs in association with two jets in proton-proton collisions at sqrts = 13 TeV},%
pdfsubject={CMS},%
pdfkeywords={CMS, physics, diboson, electroweak}
}

\maketitle
\section{Introduction}
\label{sec:Introduction}

\begin{figure*}[htb]
\centering
\includegraphics[width=0.49\textwidth]{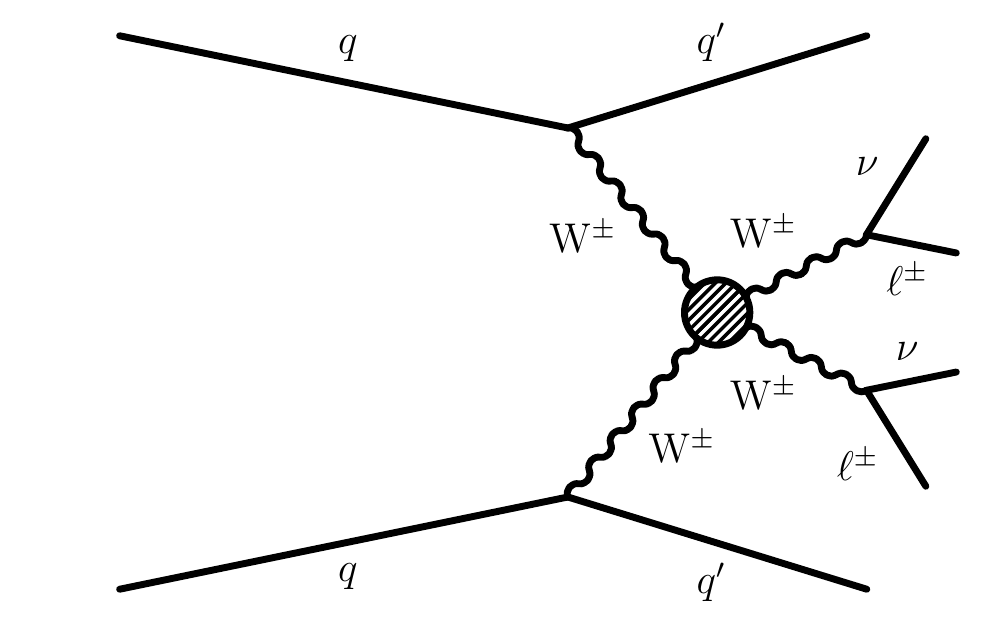}
\includegraphics[width=0.49\textwidth]{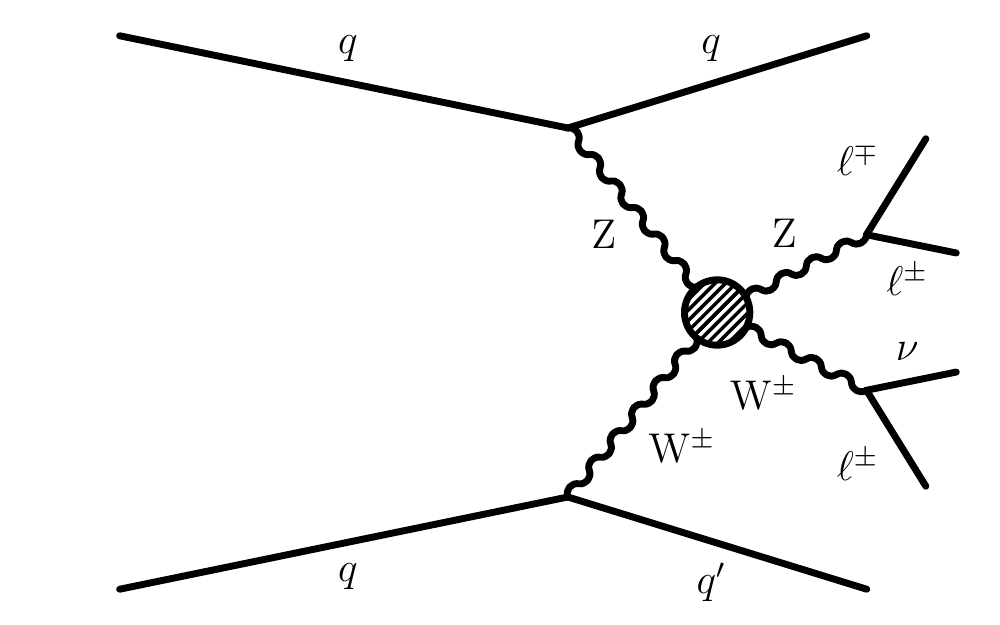}
\caption{Representative Feynman diagrams of a VBS process contributing to the EW-induced production of events containing 
$\WW$ (left) and $\WZ$ (right) boson pairs decaying to leptons, 
and two forward jets. New physics (represented by a dashed circle) in the EW 
sector can modify the quartic gauge couplings.\label{fig:feynman}}
\end{figure*}

The observation of a Higgs boson with a mass of about 
125\GeV~\cite{AtlasPaperCombination,CMSPaperCombination,CMSPaperCombination2} 
established that the $\PW$ and $\cPZ$ gauge bosons acquire mass via the Brout--Englert--Higgs 
mechanism~\cite{PhysRevLett.13.321,Higgs:1964ia,PhysRevLett.13.508,PhysRevLett.13.585,PhysRev.145.1156,PhysRev.155.1554}. 
Further insight into the electroweak (EW) symmetry breaking 
mechanism can be achieved through measurements of vector boson scattering 
(VBS) processes~\cite{Espriu:2012ih,Chang:2013aya}. At the CERN LHC interactions from VBS are characterized by 
the presence of two gauge bosons, in association with two forward jets with 
large dijet invariant mass and large rapidity separation, as shown in Fig.~\ref{fig:feynman}. 
They are part of a class of processes contributing to
diboson plus two jets production that proceeds via the EW interaction, referred to as EW-induced diboson production, at tree level,
$\mathcal{O}(\alpha^4)$, where $\alpha$ is the EW coupling. 
An additional contribution to the diboson states arises via quantum chromodynamics
(QCD) radiation of partons from an incoming quark or gluon, 
leading to tree-level contributions at $\mathcal{O}(\alpha^2\alpS^2)$, where $\alpS$ is the strong coupling. 
This class of processes is referred to as QCD-induced diboson production. 

Modifications of the VBS production cross sections are predicted in models of physics beyond the 
standard model (SM), for example through changes to the Higgs boson 
couplings to gauge bosons~\cite{Espriu:2012ih,Chang:2013aya}. 
In addition, the non-Abelian gauge structure of the EW sector of the SM 
predicts self-interactions between gauge bosons through triple and quartic 
gauge couplings, which can be probed via measurements of VBS processes~\cite{Lee:1977yc,Lee:1977eg}. 
The possible presence of anomalous quartic gauge couplings (aQGC) could result 
in an excess of events with respect to the SM predictions~\cite{Eboli:2006wa}.

This letter presents a study of VBS in $\WW$ and $\WZ$ 
channels using proton-proton ($\Pp\Pp$) collisions at $\sqrt{s}=13\TeV$. 
For the $\PW\PW$ measurement, the same-sign $\WW$ channel is chosen 
because of the smaller background yield from SM processes compared to $\PW^\pm\PW^\mp$. 
The data sample corresponds to an integrated luminosity of 
$137 \pm 2\fbinv$~\cite{CMS-PAS-LUM-17-001,CMS-PAS-LUM-17-004,CMS-PAS-LUM-18-002} 
collected with the CMS detector~\cite{Chatrchyan:2008aa} in three separate 
LHC operating periods during 2016, 2017, and 2018. The three data sets are 
analyzed independently, with appropriate calibrations and corrections, 
to account for the various LHC running conditions and the performance of the CMS detector.

\begin{figure*}[htb]
\centering
\includegraphics[width=0.49\textwidth]{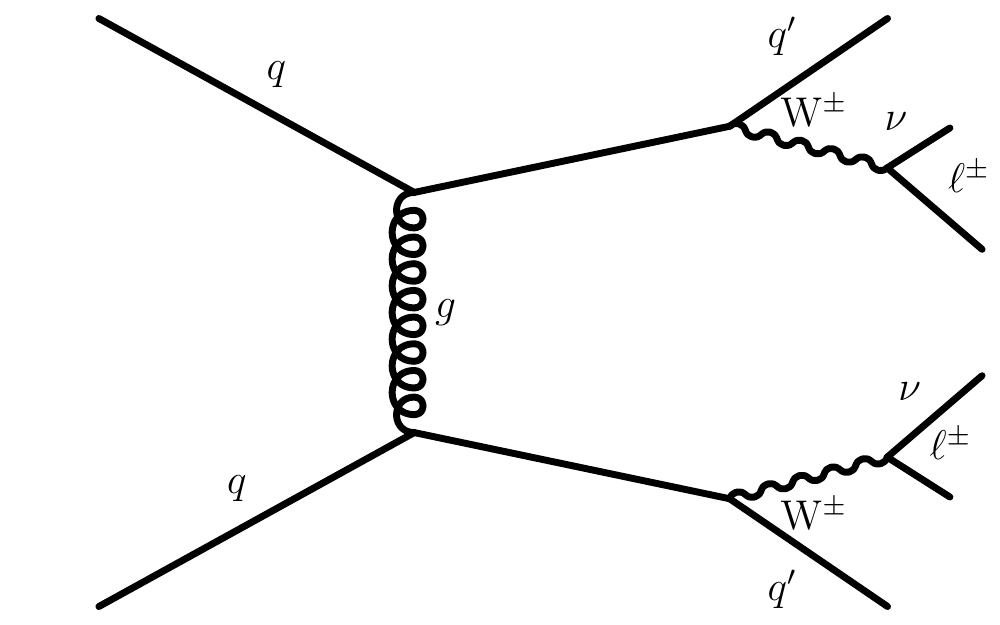}
\includegraphics[width=0.49\textwidth]{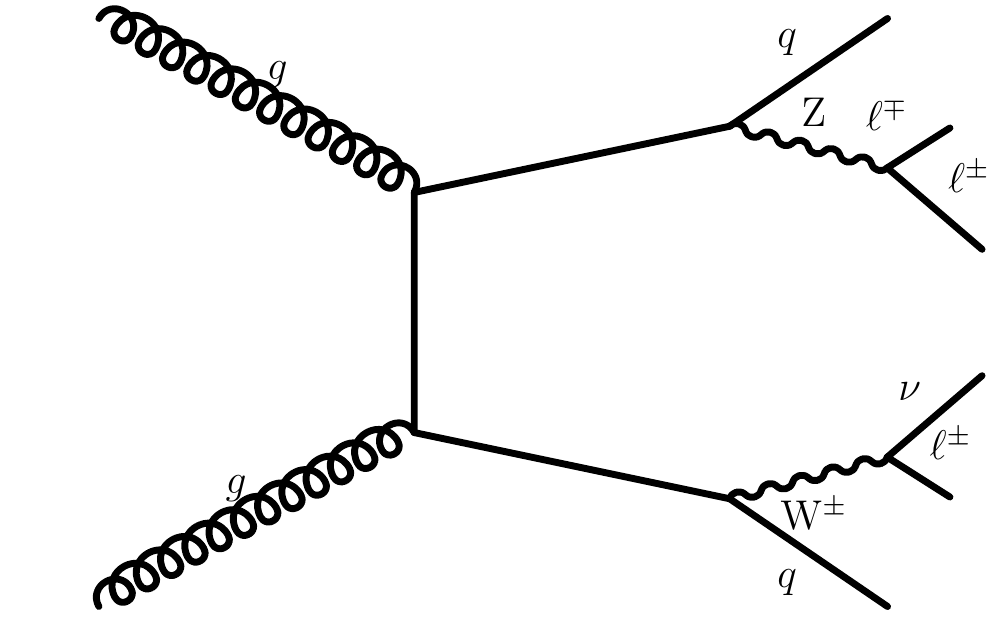}
\caption{Representative Feynman diagrams of the QCD-induced production of $\WW$ (left) and $\WZ$ (right) boson pairs decaying 
to leptons, and two jets.\label{fig:feynman_qcd}}
\end{figure*}

The measurements are performed in the leptonic decay 
modes $\WW \to \ell^\pm\PGn\ell'^\pm\PGn$ and 
$\PW^\pm\cPZ\to\ell^\pm\PGn\ell'^\pm\ell'^\mp$, where $\ell, \ell' = \Pe$, 
$\PGm$. Figure~\ref{fig:feynman} shows representative Feynman diagrams 
involving quartic vertices. Candidate events contain either two identified 
leptons of the same charge or three identified charged leptons with the total charge of $\pm$1, moderate missing transverse 
momentum ($\ptmiss$), and two jets with a large rapidity separation and a large dijet 
mass. The requirements on the dijet mass and rapidity separation reduce the contribution from the QCD-induced 
production of boson pairs in association with two jets, making 
the experimental signature an ideal topology for VBS studies. Figure~\ref{fig:feynman_qcd} 
shows representative Feynman diagrams of the QCD-induced production. The EW $\WW$ and $\WZ$ 
production cross sections are simultaneously measured by performing a binned maximum-likelihood 
fit of several distributions sensitive to these processes.

The EW production of $\WW$ at the LHC in the leptonic decay modes 
has been previously measured at $\sqrt{s}=8$ and 
13\TeV~\cite{Khachatryan:2014sta,Aad:2014zda,Sirunyan:2017ret,Aaboud:2019nmv}. 
The ATLAS and CMS Collaborations reported observations of the EW 
$\WW$ production at 13\TeV with a significance greater than 
5 standard deviations using the data collected in 2016, corresponding to 
integrated luminosities of approximately 36\fbinv. The EW 
$\WZ$ production in the fully leptonic decay modes has been studied at 8 
and 13\TeV~\cite{Aad:2016ett,Sirunyan:2019ksz,Aaboud:2018ddq}; 
the ATLAS Collaboration reported an observation at 13\TeV with a significance greater than 5 standard deviations. 
The EW production of $\WW$ and $\WZ$ 
boson pairs has also been studied in semileptonic final 
states~\cite{Aad:2019xxo}. Limits on aQGCs were also reported in Refs.~\cite{Aaboud:2016uuk,Sirunyan:2019der}.

\section{The CMS detector}
\label{sec:cms}

The central feature of the CMS apparatus is a superconducting solenoid of 
6\unit{m} internal diameter, providing a magnetic field of 3.8\unit{T}. Within 
the solenoid volume are a silicon pixel and strip tracker, a lead-tungstate 
crystal electromagnetic calorimeter (ECAL), and a brass and scintillator hadron 
calorimeter, each composed of a barrel and two endcap sections. Forward 
calorimeters extend the pseudorapidity ($\eta$) coverage provided by the barrel 
and endcap detectors up to $\abs{\eta}<5$. Muons are detected in gas-ionization chambers embedded in 
the steel magnetic flux-return yoke outside the solenoid. A more detailed description 
of the CMS detector, together with a definition of the coordinate system and 
the relevant kinematic variables, is reported in Ref.~\cite{Chatrchyan:2008aa}. 
Events of interest are selected using a two-tiered trigger 
system~\cite{Khachatryan:2016bia}. The first level, composed of custom hardware 
processors, uses information from the calorimeters and muon detectors to select events 
at a rate of around 100\unit{kHz} with a latency of 4\mus. 
The second level, known as the high-level trigger, consists of a farm of 
processors running a version of the full event reconstruction software optimized 
for fast processing, and reduces the event rate to around 1\unit{kHz} before data storage.

\section{Signal and background simulation}
\label{sec:samples}

Multiple Monte Carlo (MC) event generators are used to simulate the signal and background contributions. 
Three sets of simulated events for each process are needed to match the data-taking conditions in the various years. 

The SM EW $\WW$ and $\WZ$ processes, where both bosons decay 
leptonically, are simulated using \MGvATNLO 2.4.2~\cite{Alwall:2014hca,Madgraph,Frederix2012} at leading order (LO) 
accuracy with six EW ($\mathcal{O}(\alpha^6)$) and zero QCD vertices. 
\MGvATNLO 2.4.2 is also used to simulate the QCD-induced $\WW$ process. 
Contributions with an initial-state {\cPqb} quark are excluded from the EW $\WZ$ 
simulation because they are considered part of the $\tZq$ background process. 
Triboson processes, where the $\WZ$ boson pair is accompanied by a third vector 
boson that decays into jets, are included in the simulation.
The simulation of the aQGC processes uses the \MGvATNLO generator and employs matrix element reweighting to 
obtain a finely spaced grid of parameters for each of the probed anomalous 
couplings~\cite{Artoisenet:2010cn}. The QCD-induced $\WZ$ process is simulated at LO with up to three 
additional partons in the matrix element calculations using the \MGvATNLO generator with at least one QCD vertex at tree level. 
The different jet multiplicities are merged using the MLM scheme~\cite{MLMmerging} 
to match matrix element and parton shower jets, and the inclusive contribution is normalized to next-to-next-to-leading order (NNLO) predictions~\cite{Grazzini:2016swo}.
The interference between the EW and QCD diagrams is also produced with \MGvATNLO. 
The contribution of the interference is considered to be part of the EW production, 
leading to an increase of about 4 and 1\% of the expected yields of the 
EW $\WW$ and $\WZ$ processes in the fiducial region, respectively.

A complete set of next-to-leading order (NLO) QCD and EW corrections for the leptonic $\WW$ 
scattering process have been computed~\cite{Biedermann:2016yds,Biedermann:2017bss} and they 
reduce the LO cross section of 
the EW $\WW$ process at the level of 10-15\%, with the 
correction increasing in magnitude with increasing dilepton and dijet 
invariant masses. Similarly, the NLO QCD and EW corrections for the leptonic 
$\WZ$ scattering process have been computed at the orders of $\mathcal{O}(\alpS\alpha^6)$ 
and $\mathcal{O}(\alpha^7)$~\cite{Denner:2019tmn}, reducing the cross sections 
for the EW $\WZ$ process at the level of 10\%. The predictions for the 
cross sections of the EW $\WW$ and $\WZ$ processes are also made after applying these $\mathcal{O}(\alpS\alpha^6)$ and 
$\mathcal{O}(\alpha^7)$ corrections to \MGvATNLO LO cross sections. These 
corrections have approximately 1\% effect on the measurements and are not 
included at the data analysis level. Satisfactory agreement between 
predictions from \MGvATNLO and various event generators and fixed-order 
calculations in the fiducial region is reported in Ref.~\cite{Ballestrero:2018anz}.

The $\POWHEG$ v2~\cite{Frixione:2002ik,Nason:2004rx,Frixione:2007vw,Alioli:2008gx,Alioli:2010xd} 
generator is used to simulate the $\ttbar$, $\PQt\PW$, and other diboson processes at NLO accuracy in QCD. 
Production of $\ttbar\PW$, $\ttbar\cPZ$, $\ttbar\gamma$, and 
triple vector boson ($\PV\PV\PV$) background events is simulated at NLO accuracy 
in QCD using the \MGvATNLO 2.2.2 (2.4.2) generator~\cite{Alwall:2014hca,Madgraph,Frederix2012} for 2016 (2017 and 2018) 
samples. The $\tZq$ process is simulated at NLO in the four-flavor scheme using \MGvATNLO 2.3.3. 
The MC simulation is normalized using a cross section computed at NLO with \MGvATNLO in the 
five-flavor scheme, following the procedure of Ref.~\cite{Sirunyan:2017nbr}. 
The double parton scattering $\WW$ production is generated at LO using 
$\PYTHIA$ 8.226 (8.230)~\cite{Sjostrand:2014zea} in 2016 (2017 and 2018).

The NNPDF 3.0 NLO \cite{Ball:2014uwa} 
(NNPDF 3.1 NNLO~\cite{Ball:2017nwa}) parton distribution functions 
(PDFs) are used for simulating all 2016 (2017 and 2018) samples. 
For all processes, the parton showering and hadronization are simulated using 
$\PYTHIA$ 8.226 (8.230) in 2016 (2017 and 2018).
The modeling of the underlying event is generated using the 
CUETP8M1~\cite{Skands:2014pea,Khachatryan:2015pea} (CP5~\cite{Sirunyan:2019dfx}) tune 
for simulated samples corresponding to the 2016 (2017 and 2018) data.

All MC generated events are processed through a simulation of the CMS detector based on 
\GEANTfour~\cite{Geant} and are reconstructed with the same algorithms used for data. 
Additional $\Pp\Pp$ interactions in the same and nearby bunch 
crossings, referred to as pileup, are also simulated. 
The distribution of the number of pileup interactions in the simulation is adjusted 
to match the one observed in the data. 
The average number of pileup interactions was 23 (32) in 2016 (2017 and 2018).

\section{Event reconstruction}
\label{sec:objects}

The CMS particle-flow (PF) algorithm~\cite{Sirunyan:2017ulk} is used to combine 
the information from all subdetectors for particle reconstruction and identification. 
The vector $\ptvecmiss$ is 
defined as the projection onto the plane perpendicular to the beam axis of the negative 
vector momentum sum of all reconstructed PF objects in an event. 
Its magnitude is referred to as $\ptmiss$. 

Jets are reconstructed by clustering PF candidates using the anti-\kt 
algorithm~\cite{Cacciari:2008gp} with a distance parameter R = 0.4. 
Jets are calibrated in the simulation, and separately in data, accounting for energy 
deposits of neutral particles from pileup and any nonlinear detector response~\cite{Khachatryan:2016kdb,CMS-DP-2020-019}. 
Jets with transverse momentum $\pt>50\GeV$ and $\abs{\eta}<4.7$ are included in the analysis. 
The effect of pileup is mitigated through a charged-hadron subtraction technique, 
which removes the energy of charged hadrons not originating from the event primary vertex (PV)~\cite{Sirunyan:2020foa}. 
Jet energy corrections to the detector measurements are propagated to $\ptmiss$~\cite{Sirunyan:2019kia}. 
The PV is defined as the vertex with the largest value of summed 
physics-object $\pt^2$. Here, the physics objects are the jets clustered using the jet finding 
algorithm~\cite{Cacciari:2008gp,Cacciari:2011ma} with the tracks assigned to the 
vertex as inputs, and the associated $\ptmiss$, taken as the 
negative vector $\pt$ sum of those jets.

The \textsc{DeepCSV} \PQb\ tagging algorithm~\cite{Sirunyan:2017ezt} is used to identify events 
containing a jet that is consistent with the fragmentation of a bottom quark. 
This tagging algorithm, an improved version of previous taggers, was developed using a deep neural 
network with a more sophisticated architecture  and it provides a simultaneous training in both 
secondary vertex categories and jet flavors. For the chosen working point, the efficiency to select 
\PQb\ quark jets is about  72\% and the rate for incorrectly tagging jets 
originating from the hadronization of gluons or $\cPqu$, $\cPqd$, $\cPqs$ quarks 
is about 1\%.

Electrons and muons are reconstructed by associating a track reconstructed 
in the tracking detectors with either a cluster of energy in the 
ECAL~\cite{Khachatryan:2015hwa} or a track in the muon system~\cite{Sirunyan:2018fpa}. 
Electrons (muons) must pass ``loose'' identification criteria with $\pt>10\GeV$ and $\abs{\eta}<2.5$ (2.4) to be selected for the analysis. 
At the final stage of the lepton selection, tight working points, following the definitions provided in Refs.~\cite{Khachatryan:2015hwa,Sirunyan:2018fpa}, 
are chosen for the identification criteria, including requirements on the impact parameter 
of the candidates with respect to the PV and their isolation with 
respect to other particles in the event~\cite{Sirunyan:2018egh}. 
For electrons, the background contribution coming from a mismeasurement of the track charge is not negligible.
The sign of this charge is evaluated with three different observables that measure the electron curvature using different methods; requiring all three charge evaluations to 
agree reduces this background contribution by a factor of five with an efficiency of about 97\%~\cite{Khachatryan:2015hwa}. 
For muons, the charge mismeasurement is negligible~\cite{Chatrchyan:2009ae,Sirunyan:2019yvv}.  

\section{Event selection}
\label{sec:selection}

Collision events are collected using single-electron and single-muon
triggers that require the presence of an isolated lepton with $\pt$ larger than 
27 and 24\GeV, respectively. In addition, a set of dilepton triggers with lower $\pt$ thresholds 
are used, ensuring a trigger efficiency above 99\% for events that satisfy the subsequent offline selection.

Several selection requirements are used to isolate the VBS topology by reducing 
the contributions from background processes. By inverting some of these selection 
requirements we can select background-enriched control regions (CRs). 
In the offline analysis, events with two or three isolated charged leptons with 
$\pt>10\GeV$ and at least two jets with $\pt^\jet>50\GeV$ 
and $\abs{\eta}<4.7$ are accepted as candidate events. Jets that are within 
$\Delta R = \sqrt{\smash[b]{(\Delta\eta)^2+(\Delta\phi)^2}}<0.4$ of one of 
the identified charged leptons are excluded. 
Candidate events with four or more charged leptons satisfying the loose identification criteria 
are rejected. 

In the $\WZ$ candidate events, one of the oppositely charged same-flavor leptons from the $\cPZ$ boson 
candidate is required to have $\pt>25\GeV$ and the other $\pt>10\GeV$ with the 
invariant mass of the dilepton pair $\mll$ satisfying $\abs{\mll - m_{\cPZ}}<15\GeV$. 
In candidate events with three same-flavor leptons, the oppositely charged lepton pair with the invariant mass 
closest to the nominal $\cPZ$ boson mass $m_{\cPZ}$~\cite{Tanabashi:2018oca} is selected 
as the $\cPZ$ boson candidate. The third lepton with $\pt>20\GeV$ is associated with 
the $\PW$ boson. In addition, the trilepton invariant mass $\mlll$ is 
required to exceed 100\GeV.   

One of the leptons in the same-sign $\WW$ candidate events is required to have 
$\pt>25\GeV$ and the other $\pt>20\GeV$. The invariant mass of the dilepton pair 
$\mll$ must be greater than 20\GeV. Candidate events in the dielectron final state 
with $\abs{\mll-m_{\cPZ}}<15\GeV$ are rejected to reduce the number of $\cPZ$ boson 
background events where the charge of one of the electron candidates is 
misidentified. 

The VBS topology is targeted by requiring a large dijet invariant mass 
$\mjj>500\GeV$ and a large pseudorapidity separation 
$\detajj>2.5$. The candidate $\WW$ ($\WZ$) events 
are also required to have $\zepmax<$0.75 (1.0), where 
\begin{equation}
z_{\ell}^{*}=\bigabs{\eta^{\ell} - \frac{\eta^{\jet_{1}} + \eta^{\jet_{2}}}{2}}/\detajj
\label{eq:zep}
\end{equation}
is the Zeppenfeld variable~\cite{Rainwater:1996ud}, $\eta^{\ell}$ is the pseudorapidity 
of a lepton, and $\eta^{\jet_{1}}$ and $\eta^{\jet_{2}}$ are the pseudorapidities of 
the two candidate VBS jets. In the case of more than two jet candidates, the two jets 
with the largest $\pt$ are selected. 

The $\ptmiss$ associated with the undetected neutrinos is required to be greater than 
30\GeV. The list of selection requirements used to define the same-sign $\WW$ and 
$\WZ$ signal regions (SRs) is summarized in Table~\ref{tab:selectioncutsSR}. 
The $\WW$ SR is dominated by the EW signal process, whereas the $\WZ$ SR has 
a very large component of the QCD $\WZ$ process, as seen in Table~\ref{tab:yield}.

\begin{table*}[htbp]
\centering
  \topcaption{Summary of the selection requirements defining the $\WW$ and $\WZ$ SRs. The 
  looser lepton $\pt$ requirement on the $\WZ$ selection refers to the trailing lepton from the $\cPZ$ boson decays. 
  The $\abs{\mll - m_{\cPZ}}$ requirement is applied to the dielectron final state only in the $\WW$ SR.\label{tab:selectioncutsSR}}
\begin{tabular} {lcc}
\hline
  Variable & $\WW$ & $\WZ$ \\
\hline
Leptons                 & 2 leptons, $\pt>25/20\GeV$ & 3 leptons, $\pt>25/10/20\GeV$ \\
$\pt^\jet$              & $>$50\GeV                  & $>$50\GeV                     \\
$\abs{\mll - m_{\cPZ}}$ & $>$15\GeV ($\Pe\Pe$)       & $<$15\GeV		     \\
$\mll$                  & $>$20\GeV                  & \NA			     \\
$\mlll$                 & \NA                        & $>$100\GeV		     \\
$\ptmiss$               & $>$30\GeV		     & $>$30\GeV		     \\
\PQb\ quark veto        & Required		     & Required 		     \\
$\zepmax$               & $<$0.75		     & $<$1.0  		             \\
$\mjj$                  & $>$500\GeV                 & $>$500\GeV		     \\
$\detajj$               & $>$2.5		     & $>$2.5  		             \\
\hline
\end{tabular}
\end{table*}

\section{Background estimation}
\label{sec:backgrounds}

A combination of methods based on CRs in data and simulation is used to 
estimate background contributions. 
Uncertainties related to the theoretical and experimental predictions are 
estimated as described in Section~\ref{sec:systematics}. 
The normalization of the $\WZ$ contribution in the $\WW$ SR is constrained by the data in the 
$\WZ$ SR, which is evaluated simultaneously for the extraction of results. 
The background contribution from charge misidentification (wrong-sign) is estimated by applying 
a data-to-simulation efficiency correction due to charge-misidentified electrons. 
The electron charge misidentification rate, estimated using Drell--Yan events, 
is about 0.01 (0.3)\% in the barrel (endcap) region~\cite{Khachatryan:2015hwa,CMS-DP-2018-017}.

The nonprompt lepton backgrounds originating from leptonic decays of heavy 
quarks, hadrons misidentified as leptons, and electrons from photon conversions 
are suppressed by the identification and isolation requirements imposed on 
electrons and muons. The remaining contribution from the nonprompt lepton 
background is estimated directly from a data sample following the technique 
described in Ref.~\cite{Khachatryan:2014sta}. This sample is selected by choosing 
events using the final selection criteria, except for one of the leptons for which 
the selection is relaxed to a looser criteria and that has failed the 
nominal selection. The yield in this sample is extrapolated to the signal region 
using the efficiencies for such loosely identified leptons to pass the standard 
lepton selection criteria. This efficiency is calculated in a sample of events 
dominated by dijet production. A normalization uncertainty of 20\% is assigned for the nonprompt lepton background to include possible differences in the composition of jets between the data sample used to derive these efficiencies and the data samples in the $\WW$ and $\WZ$ SRs~\cite{Sirunyan:2018egh}.

Three CRs are used to select nonprompt lepton, $\tZq$, and $\cPZ\cPZ$ 
background-enriched events to further estimate these processes from data. 
The $\cPZ\cPZ$ process is treated as background since the analysis selection is 
not sensitive to the EW $\cPZ\cPZ$ production. 
The nonprompt lepton CR is defined by requiring the same selection as for the 
$\WW$ SR, but with the \PQb\ quark veto requirement inverted. The selected events are enriched with the nonprompt lepton background, coming mostly from semileptonic $\ttbar$ events, and further estimates the contribution of this background process in the $\WW$ SR. 
Similarly, the $\tZq$ CR is defined by requiring the same selection as for the 
$\WZ$ SR, but with the \PQb\ quark veto requirement inverted. The selected events 
are dominated by the $\tZq$ background process. 
Finally, the $\cPZ\cPZ$ CR selects events with four leptons with the same VBS-like requirements. 
The three CRs are used to estimate the normalization of the main background processes 
from data. All other background processes are estimated from simulation after applying 
corrections to account for small differences between data and simulation.

Two sets of additional CRs are defined for the $\WW$ and $\WZ$ measurements to 
validate the predictions of the background processes.  The first CR is defined 
by requiring the same selection as for the $\WW$ SR, but with a requirement 
of $200<\mjj<500\GeV$. The second CR is defined by selecting events satisfying the requirements on the leptons, $\pt^\jet$, and $\mjj$, but with at least one of the other requirements in Table~\ref{tab:selectioncutsSR} not satisfied. Good agreement between the data and predicted yields is 
observed in all these regions.

\section{Systematic uncertainties}
\label{sec:systematics}

Multiple sources of systematic uncertainty are estimated for these measurements. 
Independent sources of uncertainty are treated as uncorrelated. The impact in 
different bins of a differential distribution is considered fully correlated for 
each source of uncertainty.

The uncertainties in the integrated luminosity measurements for
the data used in this analysis are 2.5, 2.3, and 2.5\% for the 
2016, 2017, and 2018 data 
samples~\cite{CMS-PAS-LUM-17-001,CMS-PAS-LUM-17-004,CMS-PAS-LUM-18-002}, respectively. 
They are treated as uncorrelated across the three data sets. 

The simulation of pileup events assumes a total inelastic $\Pp\Pp$ cross section of 69.2\unit{mb}, 
with an associated uncertainty of 5\%~\cite{ATLAS:2016pu,Sirunyan:2018nqx}, which has 
an impact on the expected signal and background yields of about 1\%.

Discrepancies in the lepton reconstruction and identification
efficiencies between data and simulation are corrected by applying 
scale factors to all simulation samples.
These scale factors, which depend on the $\pt$ and $\eta$ for both electrons and muons, are determined using $\cPZ \to \ell\ell$ 
events in the $\cPZ$ boson peak region that were recorded with 
independent triggers~\cite{Sirunyan:2018fpa,Khachatryan:2015hwa,Sirunyan:2019bzr}. 
The uncertainty in the determination of the trigger efficiency leads to an uncertainty 
smaller than 1\% in the expected signal yield. The lepton momentum scale uncertainty is computed by varying the 
momenta of the leptons in simulation by their uncertainties, and repeating the 
analysis selection. The resulting uncertainties in the yields are $\approx$1\% for both electrons and muons. 
These uncertainties are treated as correlated across the three data sets.

The uncertainty in the calibration of the jet energy scale (JES) directly affects 
the acceptance of the jet multiplicity requirement and the $\ptmiss$ measurement. 
These effects are estimated by shifting the JES in the simulation up and down by one standard deviation. 
The uncertainty in the JES is 2--5\%, depending on $\pt$ and $\eta$~\cite{Khachatryan:2016kdb,CMS-DP-2020-019}, 
and the impact on the expected signal and background yields is about 3\%. 
There is a larger JES uncertainty in the EW $\WZ$ cross section measurement since a multivariate analysis is used 
for the measurement, which helps discriminate against the 
background processes, but also increases the corresponding uncertainty, as seen in Table~\ref{tab:impactsssww_comb_exp}.

The \cPqb\ tagging efficiency in the simulation is corrected using scale factors determined 
from data~\cite{Sirunyan:2017ezt}. These values are estimated separately for 
correctly and incorrectly identified jets. Each set of values results in uncertainties in the \cPqb\ tagging 
efficiency of about 1--4\%, and the impact on the expected signal and background yields is about 1\%. 
The uncertainties in the JES and \cPqb\ tagging are treated as 
uncorrelated across the three data sets.

Because of the choice of the QCD renormalization and factorization scales, the theoretical uncertainties are estimated by varying these scales independently up and down by a factor of two from their nominal values 
(excluding the two extreme variations) and taking the largest cross section variations as the uncertainty~\cite{Ballestrero:2018anz}. 
The PDF uncertainties are evaluated according to the procedure described in Ref.~\cite{PDFLHC}. 
The statistical uncertainties that are associated with the limited number of simulated events and data events used to estimate the nonprompt lepton background are also considered 
as systematic uncertainties; the data events are the dominant contribution.

A summary of the relative systematic uncertainties in the EW $\WW$ and $\WZ$ 
cross sections is shown in Table~\ref{tab:impactsssww_comb_exp}. The slightly larger 
theoretical uncertainty in the EW $\WZ$ cross section measurement 
arises from the difficulty of disentangling the EW and QCD components in the discriminant fit.

\begin{table*}[htbp]
\centering
\topcaption{
Relative systematic uncertainties in the EW $\WW$ and $\WZ$ cross section measurements in units of percent.
\label{tab:impactsssww_comb_exp}}
\begin{tabular}{lcc}
\hline
Source of uncertainty           & $\WW$ (\%) & $\WZ$ (\%) \\
\hline
Integrated luminosity           &  1.5  &  1.6  \\
Lepton measurement            	&  1.8  &  2.9  \\
Jet energy scale and resolution &  1.5  &  4.3  \\
Pileup  			&  0.1  &  0.4  \\
\cPqb tagging   		&  1.0  &  1.0  \\
Nonprompt rate                  &  3.5  &  1.4  \\
Trigger                         &  1.1  &  1.1  \\
Limited sample size             &  2.6  &  3.7  \\
Theory                        	&  1.9  &  3.8  \\[\cmsTabSkip]

Total systematic uncertainty    &  5.7  &  7.9  \\[\cmsTabSkip]

Statistical uncertainty         &  8.9  & 22    \\[\cmsTabSkip]

Total uncertainty               & 11    & 23    \\
\hline
\end{tabular}
\end{table*}

\section{Results}
\label{sec:results}

To discriminate between the signals and the remaining backgrounds, 
a binned maximum-likelihood fit is performed using the $\WW$ and $\WZ$ SRs, and the 
nonprompt lepton, $\tZq$, and $\cPZ\cPZ$ CRs. The normalization factors for the 
$\tZq$ and $\cPZ\cPZ$ background processes are included in the 
maximum-likelihood fit together with the EW $\WW$, EW $\WZ$, 
and QCD $\WZ$ signal cross sections. The QCD $\WW$ contribution 
is small and is taken from the SM prediction. The systematic uncertainties are 
treated as nuisance parameters in the fit~\cite{Junk,Read}.  

The value of $\mjj$ is effective in discriminating between the signal and 
background processes because VBS topologies typically exhibit large values 
for the dijet mass. The value of $\mll$ is also effective in discriminating between 
signal and background processes because the nonprompt lepton processes tend to 
have rather small $\mll$ values. A two-dimensional distribution is used in the fit 
for the $\WW$ SR with 8 bins in $\mjj$ 
([500, 650, 800, 1000, 1200, 1500, 1800, 2300, $\infty$]\GeV) and 4 bins 
in $\mll$  ([20, 80, 140, 240, $\infty$]\GeV).

A boosted decision tree (BDT) is trained using the \textsc{tmva} package~\cite{Hocker:2007ht} 
with gradient boosting and optimized on simulated events to better separate the 
EW $\WZ$ and QCD $\WZ$ processes in the $\WZ$ SR by exploring the kinematic differences. 
Several discriminating observables are used as the BDT inputs, 
including the jet and lepton kinematics and $\ptmiss$, as listed in 
Table~\ref{tab:BDT_variables}. A larger set of discriminating observables was 
studied, but only variables improving the sensitivity and showing some 
signal-to-background separation are retained. 
The BDT score distribution is used for the $\WZ$ SR in the fit with 8 bins 
([-1, -0.28, 0.0, 0.23, 0.43, 0.60, 0.74, 0.86, 1]). The $\mjj$ 
distribution is used for the CRs in the fit with 4 bins 
([500, 800, 1200, 1800, $\infty$]\GeV). The bin boundaries are chosen to 
have the same EW $\WW$ and $\WZ$ contributions across the bins as expected from simulation.

\begin{table*}[htbp]
\centering
\topcaption{List and description of all the input variables used in the BDT analysis for the $\WZ$ SR.\label{tab:BDT_variables}}
\begin{tabular}{ l c}
\hline
Variable & Definition \\
\hline
$\mjj$                                                            & Mass of the leading and trailing jets system \\
$\detajj$	                                                  & Absolute difference in rapidity of the leading and trailing jets \\
$\dphijj$	                                                  & Absolute difference in azimuthal angles of the leading and trailing jets \\
$\pt^{\jet1}$		                                          & $\pt$ of the leading jet \\
$\pt^{\jet2}$		                                          & $\pt$ of the trailing jet \\
$\eta^{\jet1}$		                                          & Pseudorapidity of the leading jet \\
\multirow{2}{*}{$\abs{\eta^{\PW}- \eta^{\cPZ}}$}                  & Absolute difference between the rapidities of the $\cPZ$ boson \\
                                                                  &  and the charged lepton from the decay of the $\PW$ boson \\
$z_{\ell_{i}}^{*} (i = 1-3)$                                      & Zeppenfeld variable of the three selected leptons \\
$z_{3\ell}^{*}$	                                                  & Zeppenfeld variable of the vector sum of the three leptons \\
$\Delta R_{\jet1,\cPZ}$                                           & $\Delta R$ between the leading jet and the $\cPZ$ boson \\
\multirow{2}{*}{$\abs{\vec{\pt}^{\mathrm{tot}}}/\sum_i \pt^i$}    & Transverse component of the vector sum of the bosons \\
                                                                  &  and tagging jets momenta, normalized to their scalar $\pt$ sum \\
\hline
\end{tabular}
\end{table*}

\begin{figure*}[htbp]
\centering
\includegraphics[width=0.49\textwidth]{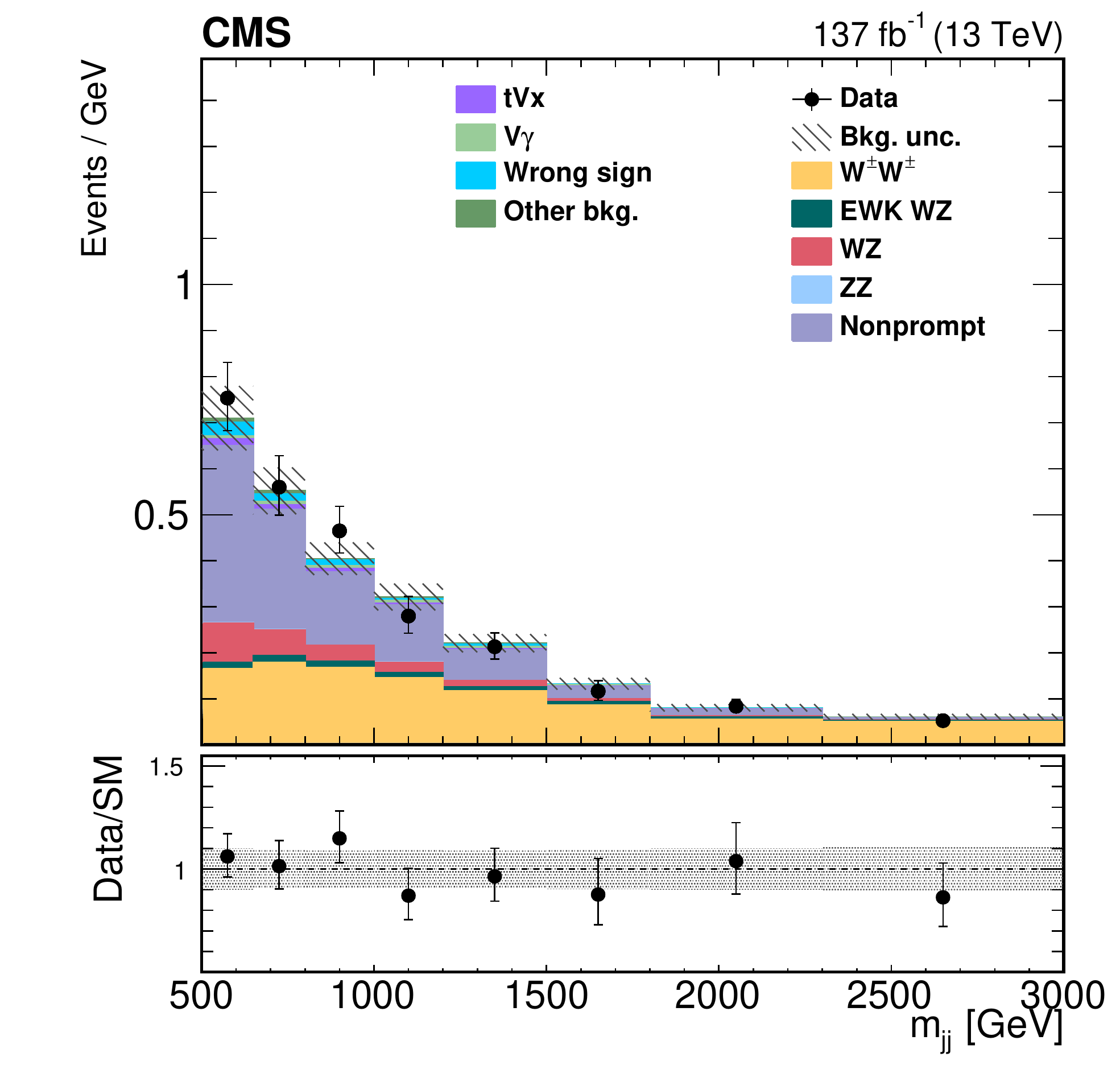}
\includegraphics[width=0.49\textwidth]{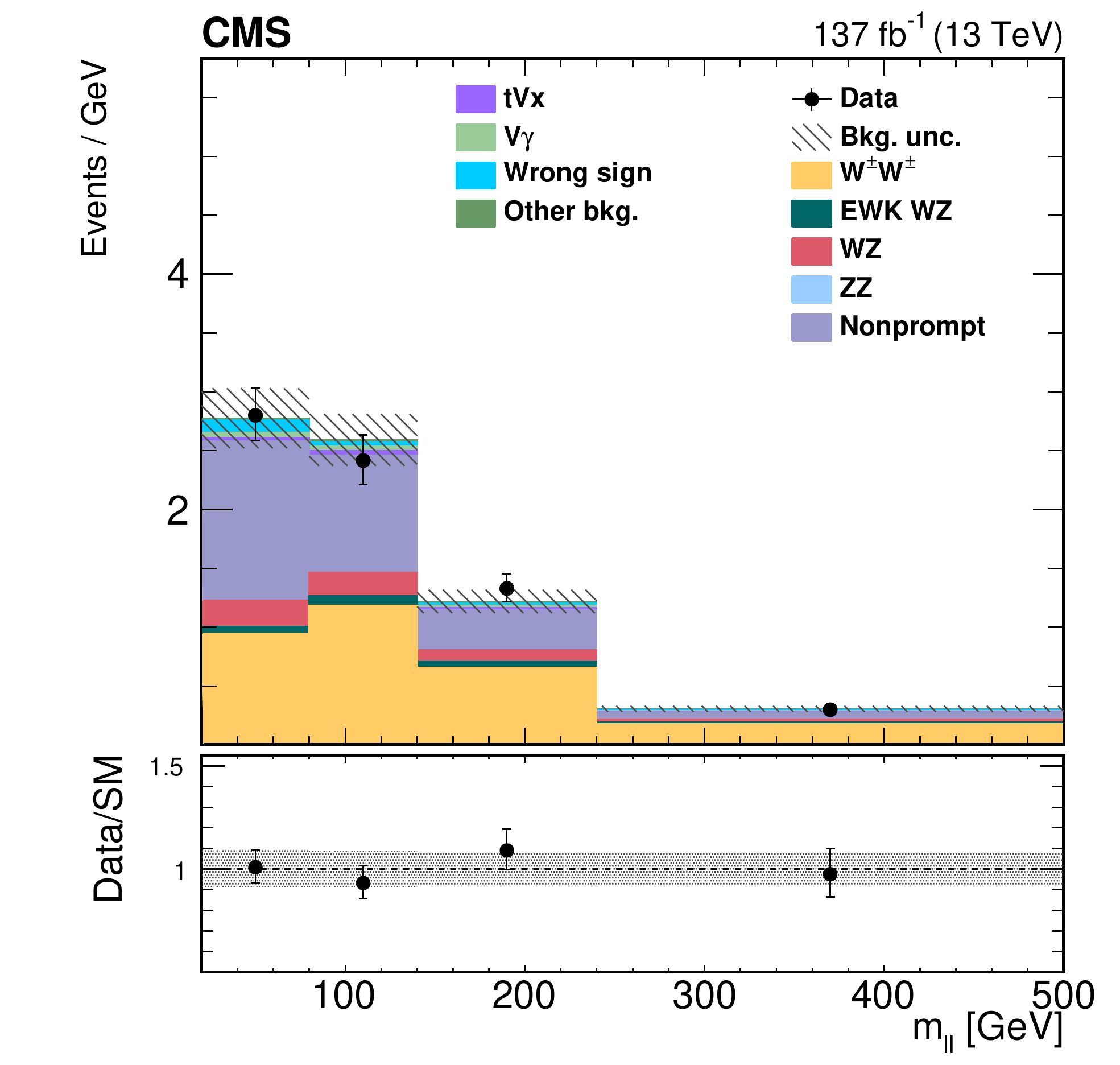}
\includegraphics[width=0.49\textwidth]{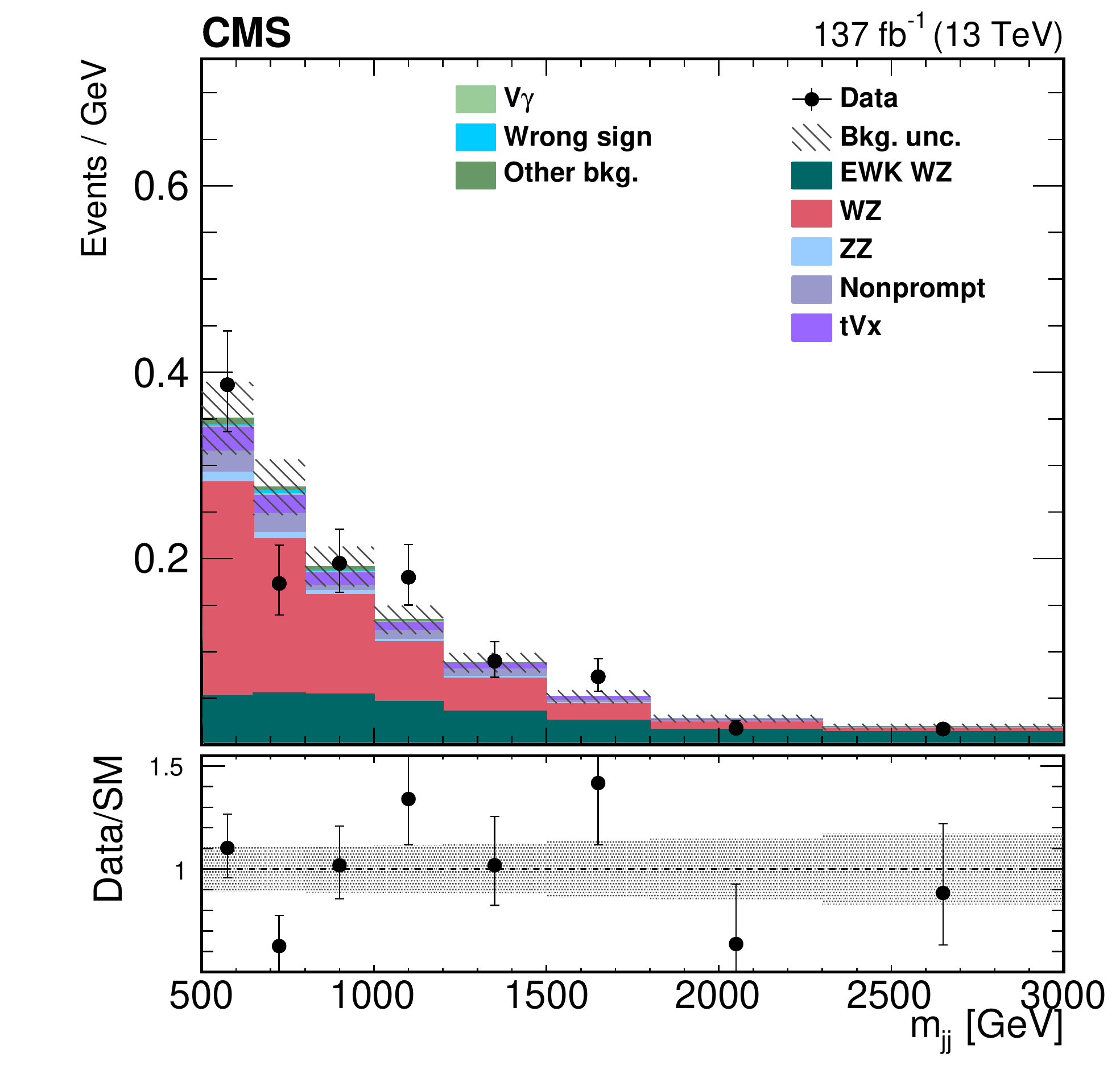}
\includegraphics[width=0.49\textwidth]{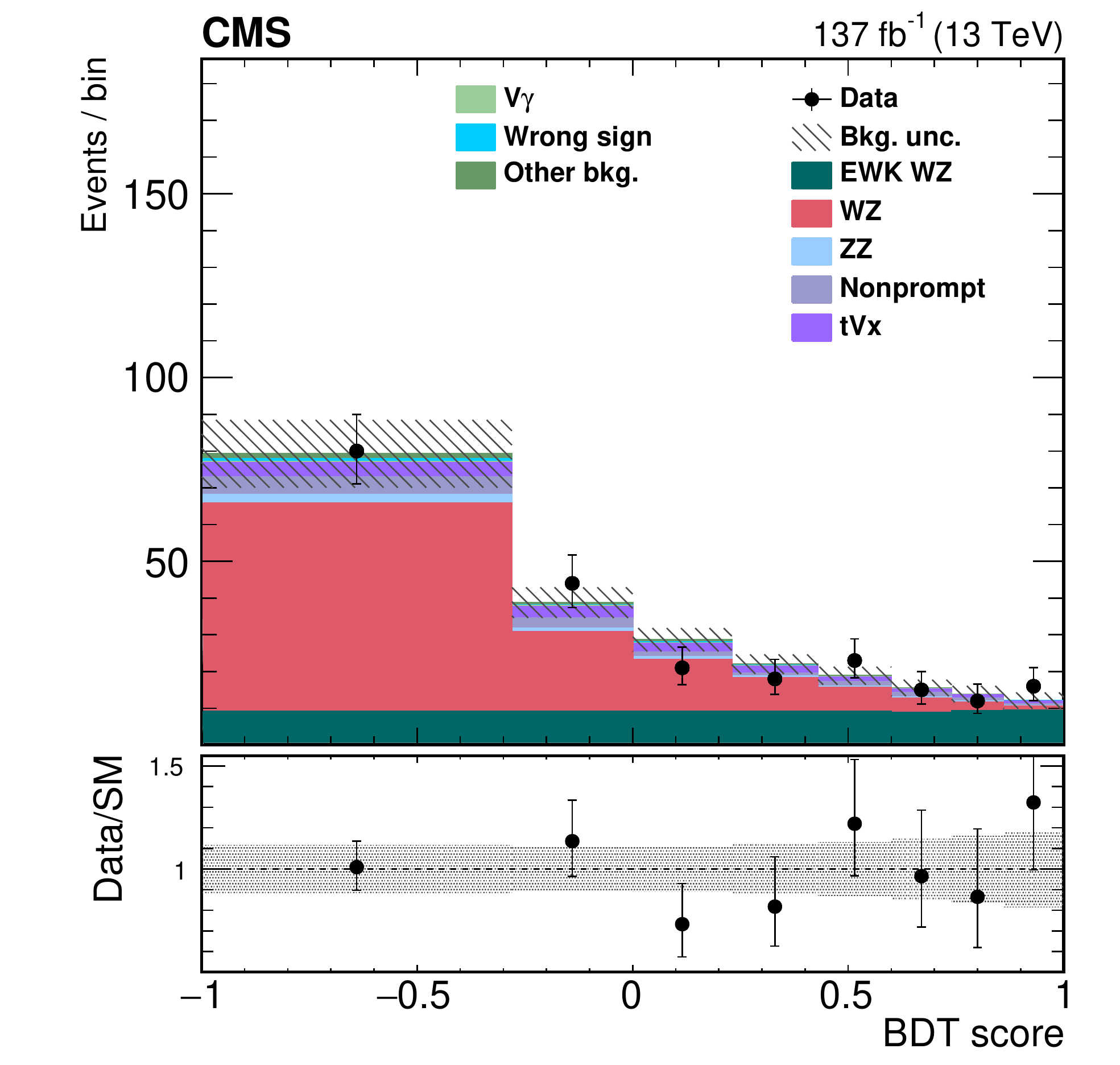}
\caption{Distributions of $\mjj$ (upper left) and $\mll$ (upper right) 
in the $\WW$ SR, and the distributions of $\mjj$ (lower left) 
and BDT score (lower right) in the $\WZ$ SR. The predicted yields are shown 
with their best fit normalizations from the simultaneous fit. 
Vertical bars on data points represent the statistical uncertainty in the data. The contribution of the QCD $\WW$ process is included together with the EW $\WW$ process. The histograms for $\tVx$ backgrounds include the contributions from $\ttbar\PV$ and $\tZq$ processes. 
The histograms for other backgrounds include the contributions from double parton scattering and $\PV\PV\PV$ processes. 
The histograms for wrong-sign background include the contributions from oppositely charged dilepton final states from $\ttbar$, $\PQt\PW$, $\PW^{+}\PW^{-}$, and Drell--Yan processes. 
The overflow is included in the last bin. The bottom panel in each figure 
shows the ratio of the number of events observed in data to that of the total SM prediction. 
The gray bands represent the uncertainties in the predicted yields.\label{fig:ssww_signalsel}}
\end{figure*}

The distributions of $\mjj$ and $\mll$ in the $\WW$ SR, and 
the distributions of $\mjj$ and BDT score in the $\WZ$ SR are shown in 
Fig.~\ref{fig:ssww_signalsel}. The data yields, together 
with the numbers of fitted signal and background events, are given in Table~\ref{tab:yield}. The table also shows the result of a fit to the Asimov data set~\cite{CLs}.
The significance of the EW $\WZ$ signal is quantified from the p-value 
using a profile ratio test statistic~\cite{Junk,Read} and asymptotic results 
for the test statistic~\cite{CLs}. The observed (expected) statistical 
significance of the EW $\WZ$ signal is 6.8 (5.3) standard deviations, 
while the statistical significance of the EW $\WW$ signal is far above 5 standard deviations.  

\begin{table*}[htbp]
\centering
\topcaption{Expected yields from SM processes and observed data events in $\WW$ 
and $\WZ$ SRs. The combination of the statistical and systematic uncertainties 
is shown. The expected yields are shown with their best fit normalizations from the simultaneous fit to the Asimov data set and to the data. The signal yields do not include the QCD and EW NLO corrections.\label{tab:yield}}
\begin{tabular}{ccccc}
\hline
Process                  &  \multicolumn{2}{c}{$\WW$ SR}     &  \multicolumn{2}{c}{$\WZ$ SR} \\
                         & Asimov data set & Data & Asimov data set & Data \\
\hline
EW $\WW$                         &   209   $\pm$   26    &   210   $\pm$   26	 &	    \NA 	 &	    \NA 	  \\
QCD $\WW$                        &    13.8 $\pm$    1.6  &    13.7 $\pm$    2.2  &	    \NA 	 &	    \NA 	  \\
Interference $\WW$               &     8.4 $\pm$    2.3  &     8.7 $\pm$    2.3  &	    \NA 	 &	    \NA 	  \\
EW $\WZ$              	         &    14.1 $\pm$    4.0  &    17.8 $\pm$    3.9  &	54   $\pm$  15   &	69   $\pm$   15   \\ 
QCD $\WZ$             	         &    43   $\pm$    6.7  &    42.7 $\pm$    7.4  &     118   $\pm$  17   &     117   $\pm$   17   \\ 
Interference $\WZ$    	         &     0.3 $\pm$    0.1  &     0.3 $\pm$    0.2  &	 2.2 $\pm$   0.9 &	 2.7 $\pm$    1.0 \\ 
$\cPZ\cPZ$                 	 &     0.7 $\pm$    0.2  &     0.7 $\pm$    0.2  &	 6.1 $\pm$   1.7 &	 6.0 $\pm$    1.8 \\ 
Nonprompt              	         &   211   $\pm$   43    &   193   $\pm$   40	 &	14.6 $\pm$   7.4 &	14.4 $\pm$    6.7 \\ 
$\tVx$                           &     7.8 $\pm$    1.9  &     7.4 $\pm$    2.2  &	15.1 $\pm$   2.7 &	14.3 $\pm$    2.8 \\ 
$\PW\gamma$             	 &     9.0 $\pm$    1.8  &     9.1 $\pm$    2.9  &	 1.1 $\pm$   0.3 &	 1.1 $\pm$    0.4 \\ 
Wrong-sign                       &    13.5 $\pm$    6.5  &    13.9 $\pm$    6.5  &	 1.6 $\pm$   0.5 &	 1.7 $\pm$    0.7 \\ 
Other background                 &     5.0 $\pm$    1.3  &     5.2 $\pm$    2.1  &	 3.3 $\pm$   0.6 &	 3.3 $\pm$    0.7 \\[\cmsTabSkip]

Total SM                         &   535   $\pm$   52    &   522   $\pm$   49	 &     216   $\pm$  21   &	229  $\pm$   23   \\[\cmsTabSkip]

Data                             &    \multicolumn{2}{c}{524}    &   \multicolumn{2}{c}{229}     \\
\hline
\end{tabular}
\end{table*}

\subsection{Inclusive and differential fiducial cross section measurements}

The fiducial region is defined by a common set of kinematic requirements in 
the muon and electron final states at the generator level, emulating the selection 
performed at the reconstruction level. The measured distributions, after 
subtracting the contributions from the background processes, are corrected for 
detector resolution effects and inefficiencies. The leptons at generator level 
are selected at the so-called dressed level by combining the four-momentum of each 
lepton after the final-state photon radiation with that of photons found within 
a cone of $\Delta R=0.1$ around the lepton. The $\WW$ fiducial region is defined by requiring two same-sign 
leptons with $\pt>20\GeV$, $\abs{\eta}<2.5$, and $\mll>20\GeV$, and two jets 
with $\mjj>500\GeV$ and $\detajj>2.5$. The jets at generator level are clustered from stable particles, excluding 
neutrinos, using the anti-$\kt$ clustering algorithm with R = 0.4, and are required 
to have $\pt>50\GeV$ and $\abs{\eta}<4.7$. The jets within $\Delta R<0.4$ of 
the selected charged leptons are not included. The $\WZ$ fiducial 
region is defined by requiring three leptons with $\pt>20\GeV$, $\abs{\eta}<2.5$, 
a pair of opposite charge same-flavor lepton pair with $\abs{\mll-m_{\cPZ}}<15\GeV$, 
and two jets with $\mjj>500\GeV$ and $\detajj>2.5$. \MGvATNLO is used to extrapolate from the reconstruction level to the fiducial 
phase space. Electrons and muons produced in the 
decay of a $\tau$ lepton are not included in the definition of the fiducial region. Nonfiducial events, $\ie$, events selected at the reconstructed level that do not 
satisfy the fiducial requirements, are included as background processes in the 
simultaneous fit.

\begin{figure*}[htbp]
\centering
\includegraphics[width=0.41\textwidth]{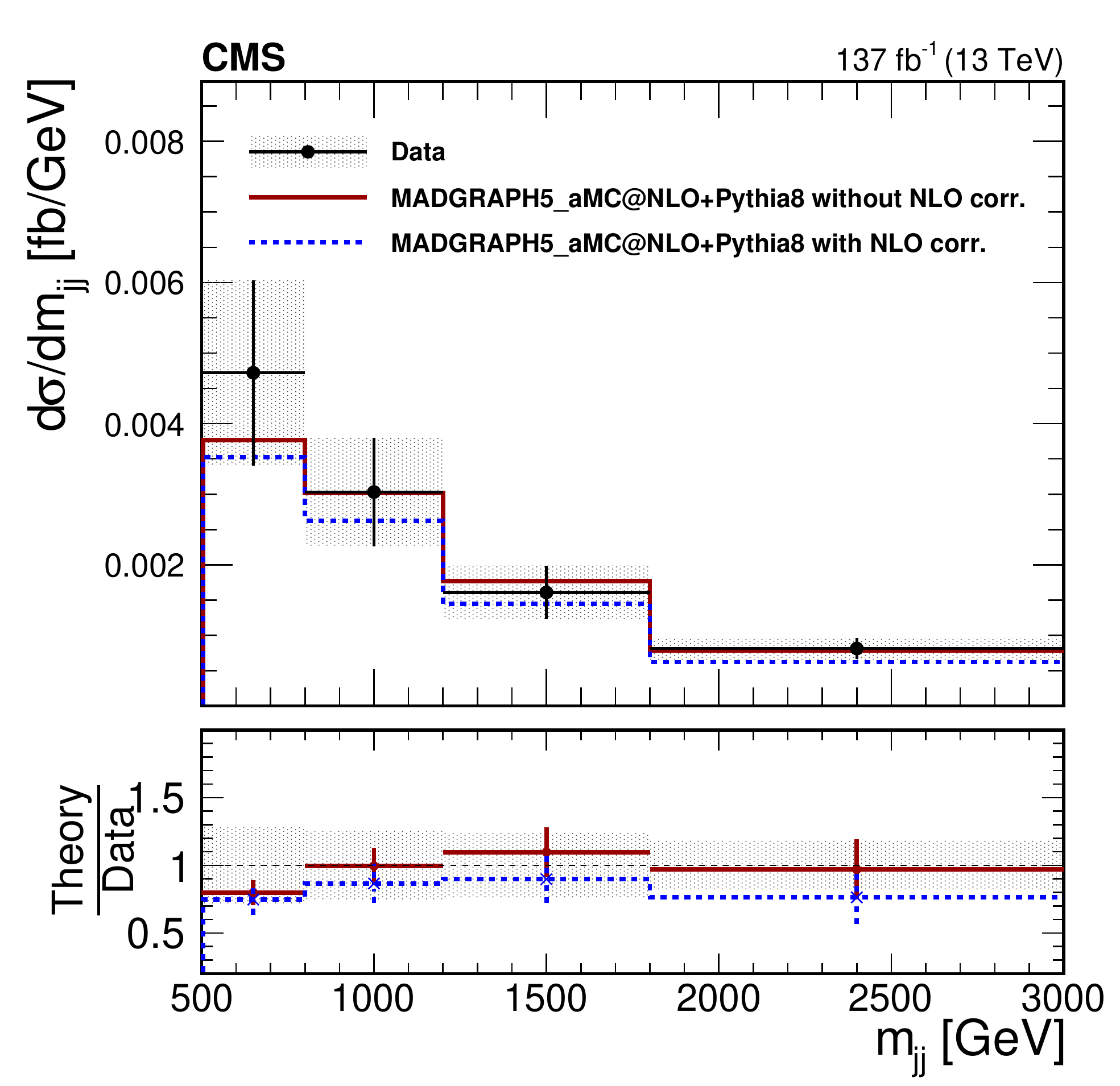}
\includegraphics[width=0.41\textwidth]{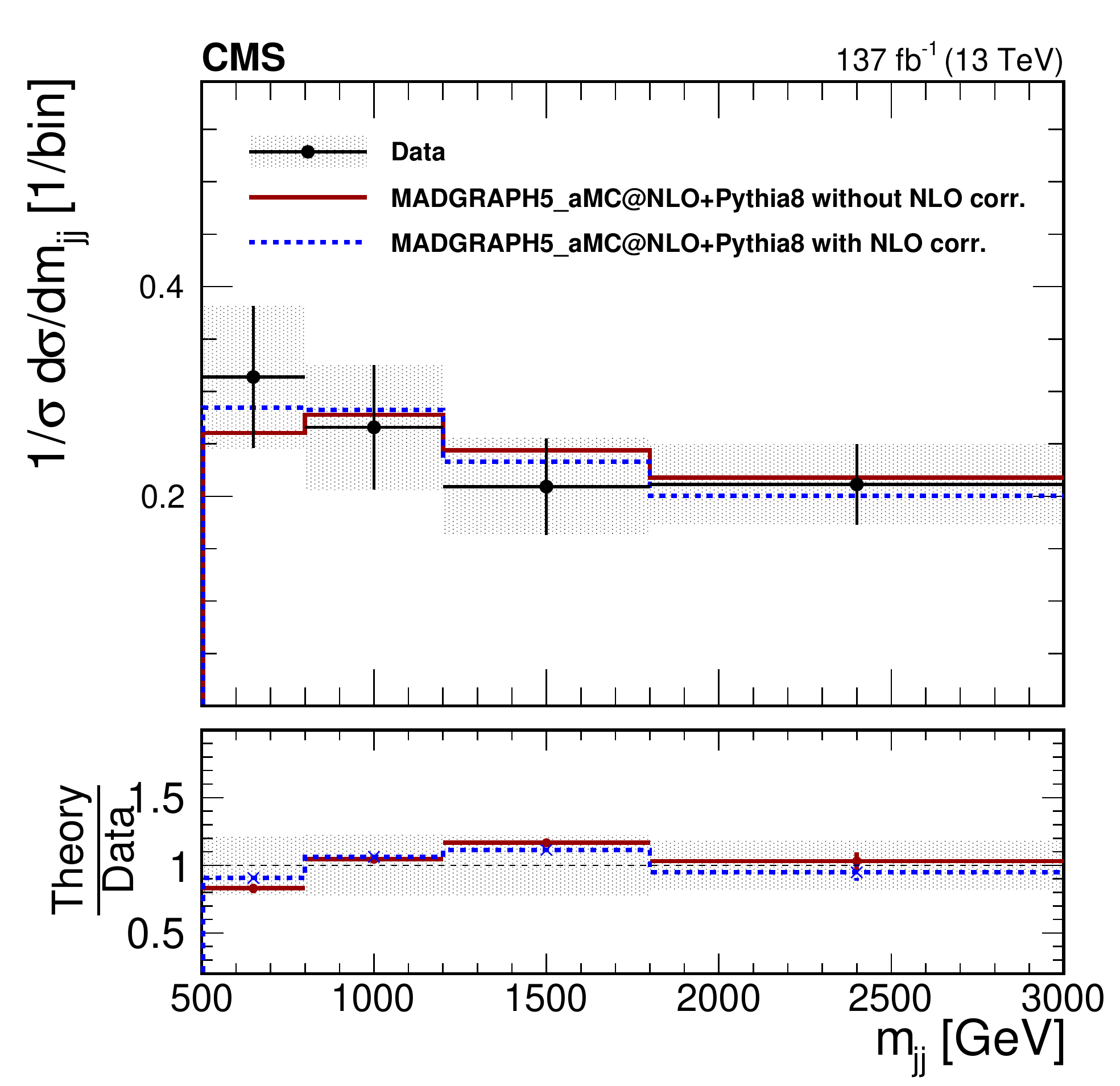}
\includegraphics[width=0.41\textwidth]{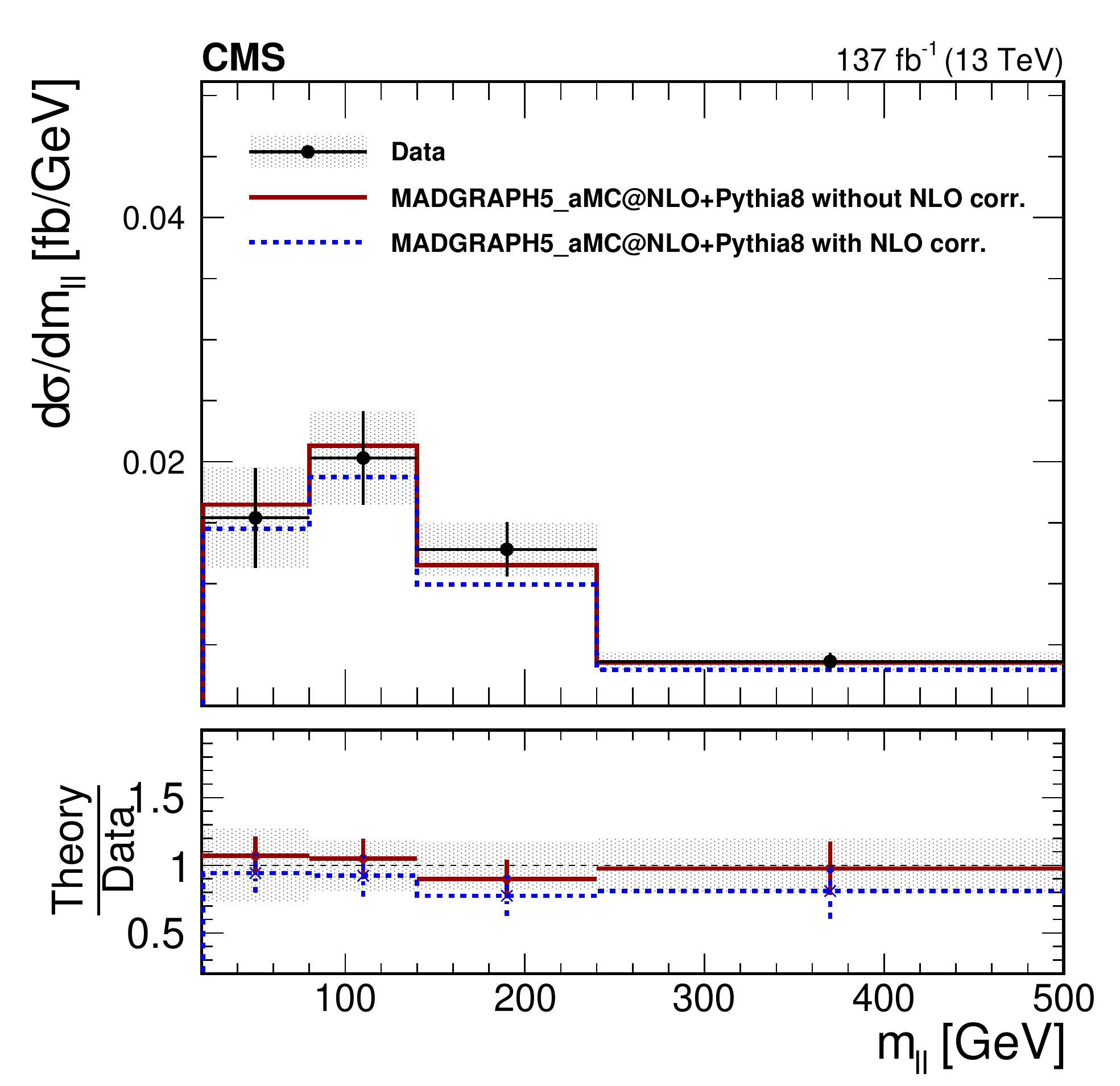}
\includegraphics[width=0.41\textwidth]{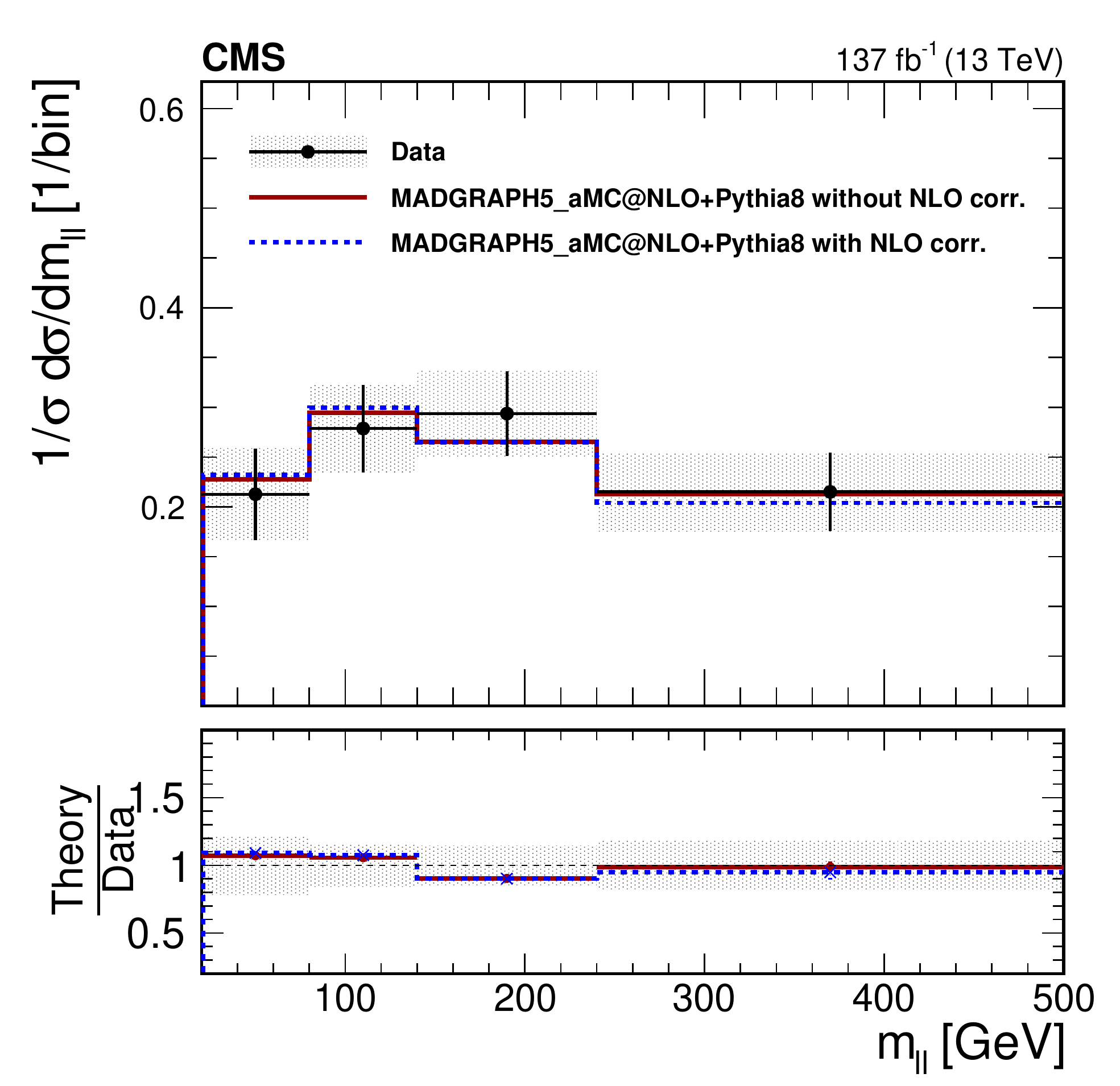}
\includegraphics[width=0.41\textwidth]{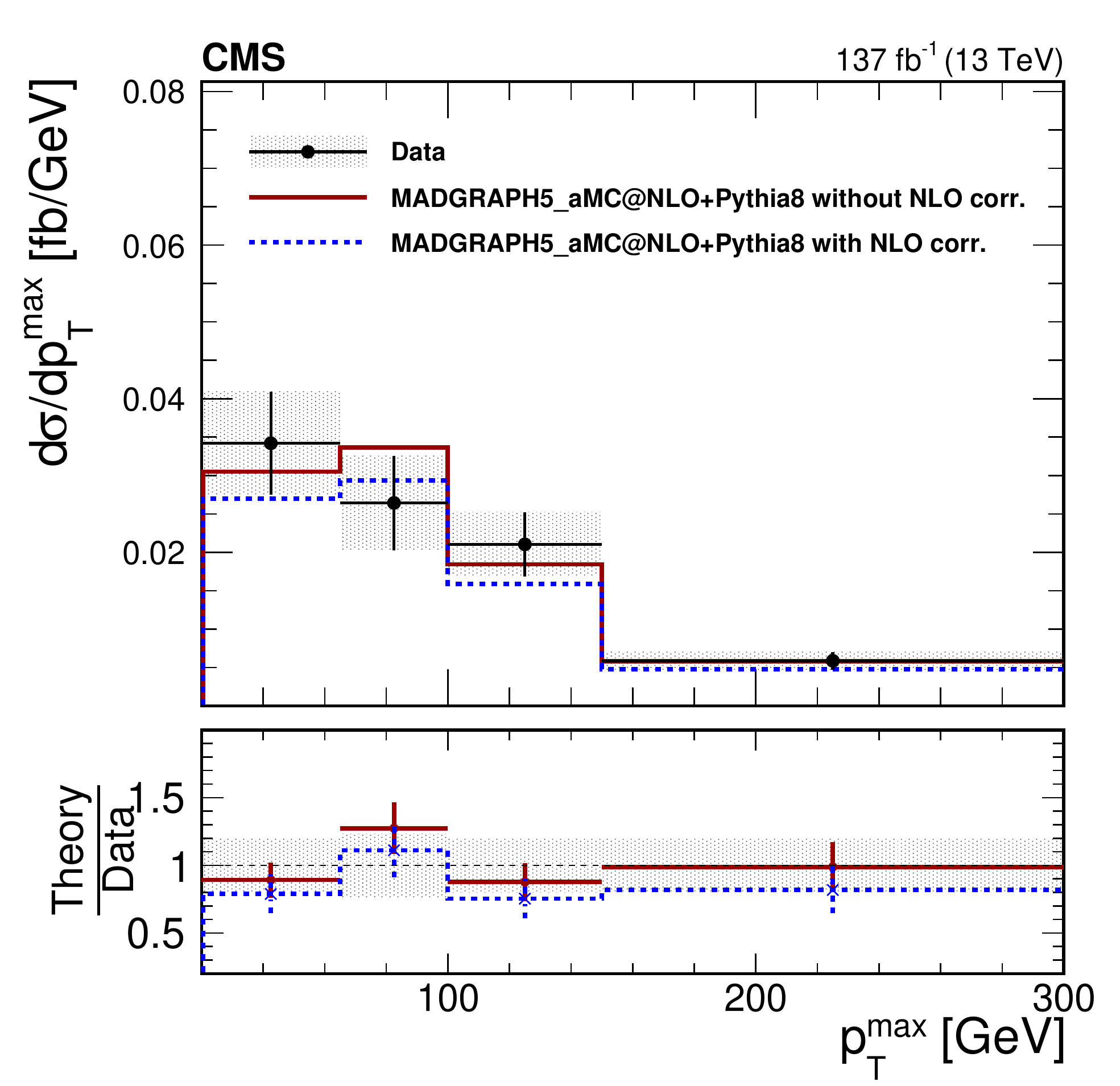}
\includegraphics[width=0.41\textwidth]{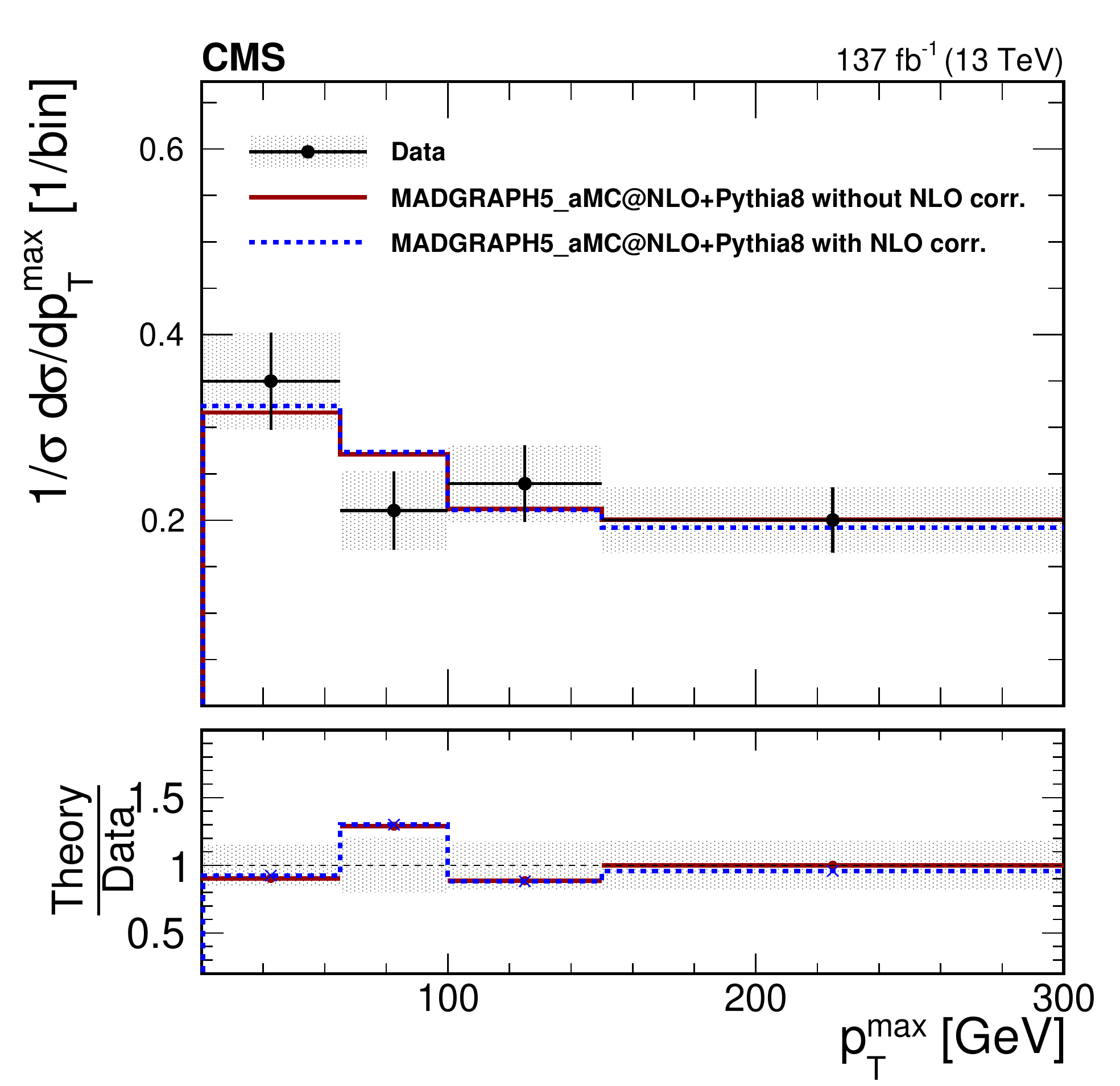}
\caption{The measured absolute (left) and normalized (right) $\WW$ cross section 
measurements in bins of $\mjj$  (upper), $\mll$ (middle), and $\ptmax$ (lower). 
The ratios of the predictions to the data are also shown. The measurements are compared with the predictions 
from \MGvATNLO at LO. The shaded bands around the data points correspond to the measurement uncertainty. 
The error bars around the predictions correspond to the combined statistical, PDF, and scale uncertainties. 
Predictions with applying the $\mathcal{O}(\alpS\alpha^6)$ and $\mathcal{O}(\alpha^7)$ corrections 
to the \MGvATNLO LO cross sections, as described in the text, are also shown (dashed blue).\label{fig:unf_ww_normalizedy}}
\end{figure*}

\begin{figure*}[htbp]
\centering
\includegraphics[width=0.49\textwidth]{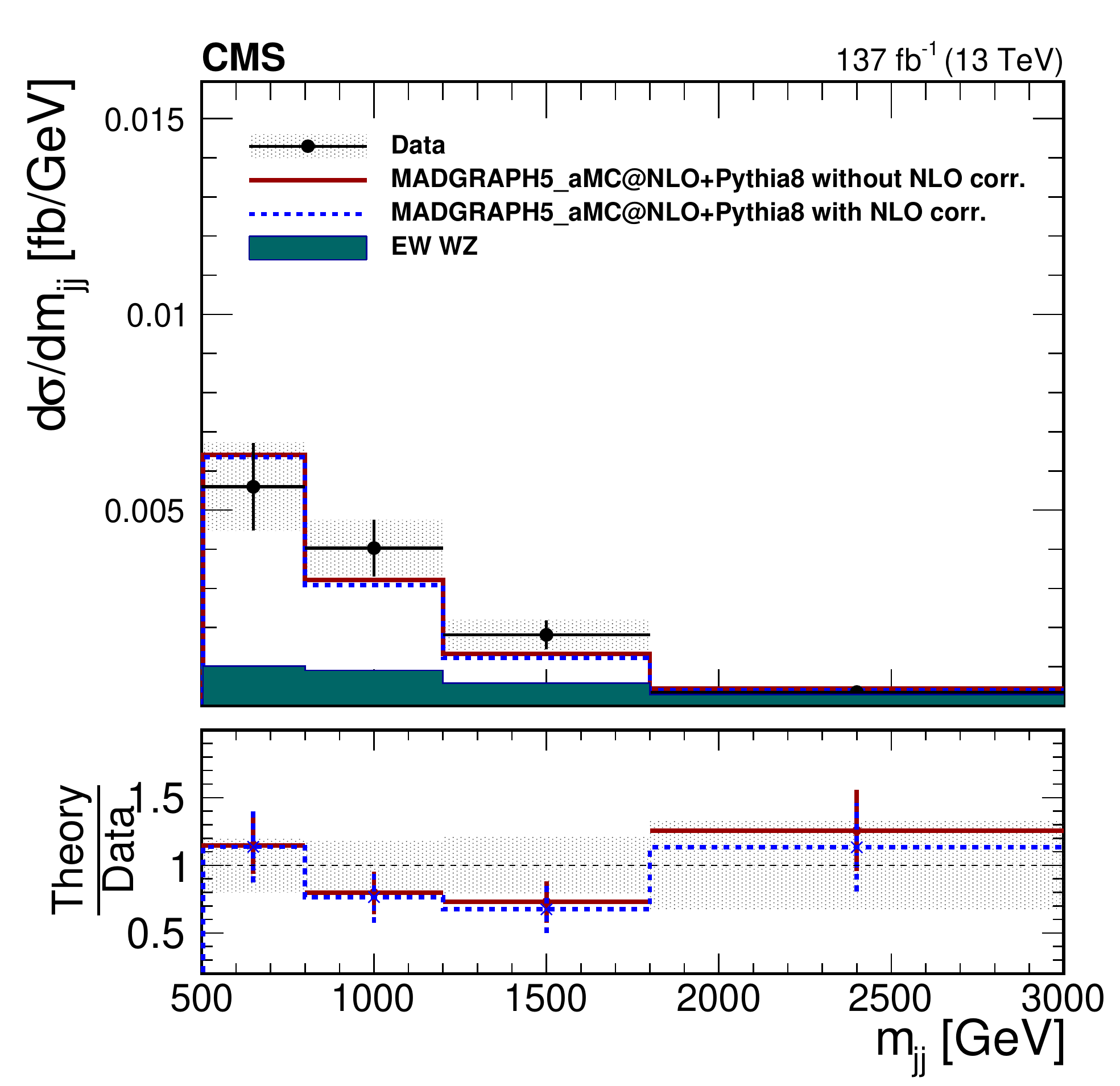}
\includegraphics[width=0.49\textwidth]{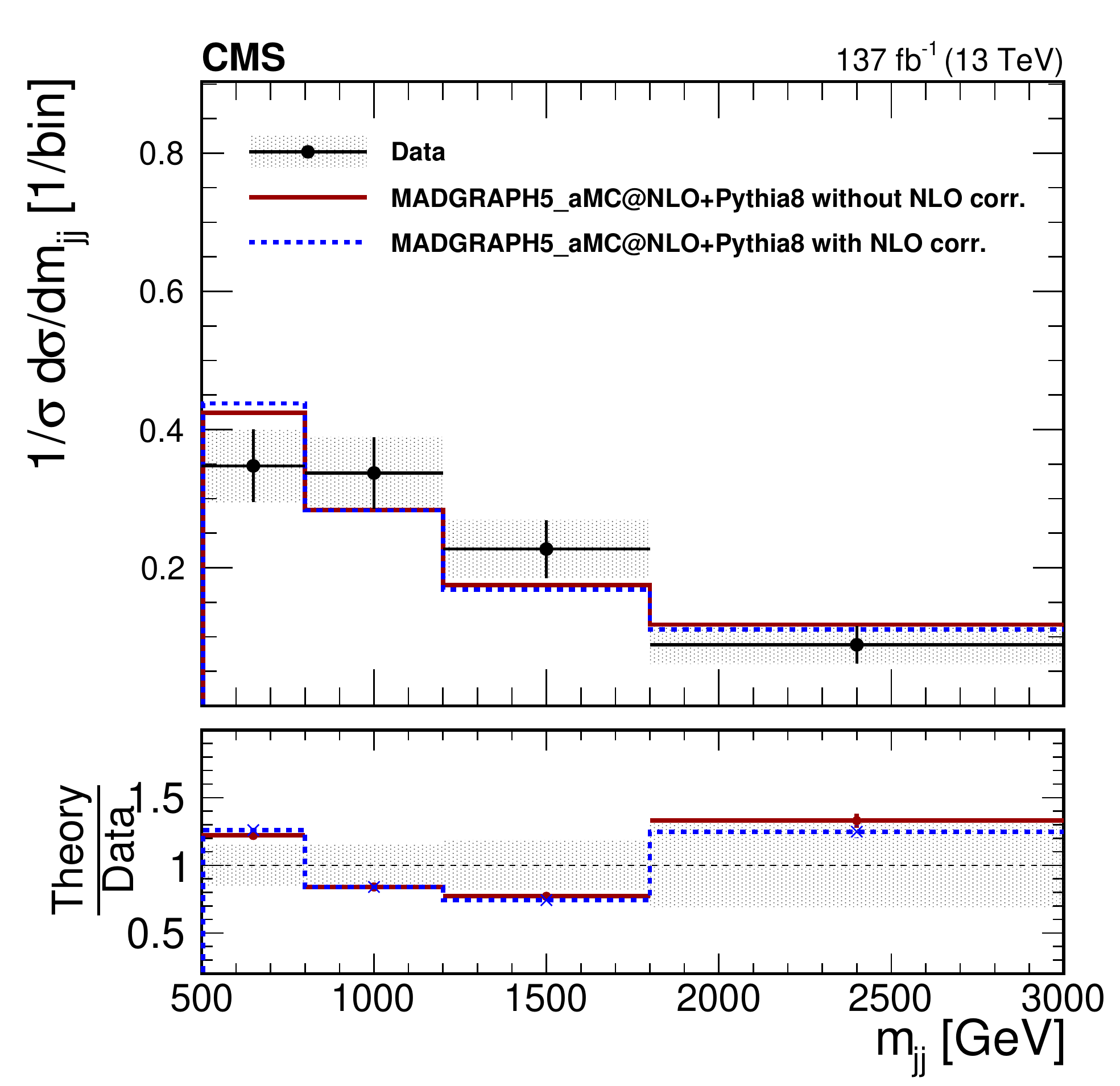}
\caption{The measured absolute (left) and normalized (right) $\WZ$ cross section measurements in 
bins of $\mjj$. The ratios of the predictions to the data are also shown. The measurements are compared with the predictions 
from \MGvATNLO at LO. The shaded bands around the data points correspond to the measurement uncertainty. 
The error bars around the predictions correspond to the combined statistical, PDF, and scale uncertainties. 
Predictions with applying the $\mathcal{O}(\alpS\alpha^6)$ and $\mathcal{O}(\alpha^7)$ corrections 
to the \MGvATNLO LO cross sections, as described in the text, are shown (dashed blue). 
The \MGvATNLO predictions in the EW total cross sections are also shown (dark cyan).\label{fig:unf_wz_normalizedy}}
\end{figure*}

Inclusive cross section measurements for the EW $\WW$, EW+QCD $\WW$, 
EW $\WZ$, QCD $\WZ$, and EW+QCD $\WZ$ processes, and the 
theoretical predictions are summarized in Table~\ref{tab:default_fid_inc}. 
To perform absolute and normalized differential production cross section measurements, 
signal templates from different bins of differential-basis observable values predicted 
by the event generator are built. Each signal template is considered 
as a separate process in the simultaneous binned maximum-likelihood fit. 
In the normalized cross section measurements, the individual 
cross sections in every fiducial region and the total production cross section are simultaneously evaluated, reducing the systematic 
uncertainties. The signal extraction at reconstruction level and the
unfolding into the  generator level bins are performed in a single step in the simultaneous fit. The bin migration effects due to the detector resolution are negligible. The measurement is compared with the \MGvATNLO predictions at LO. The \MGvATNLO predictions including 
the $\mathcal{O}(\alpS\alpha^6)$ and $\mathcal{O}(\alpha^7)$ corrections in the 
EW $\WW$ and $\WZ$ processes are also included in Table~\ref{tab:default_fid_inc}. The measured absolute 
and normalized $\WW$ differential cross sections in bins of 
$\mjj$, $\mll$, and leading lepton $\pt$ ($\ptmax$) are shown in 
Fig.~\ref{fig:unf_ww_normalizedy}. 
The absolute cross sections are shown in fb per \GeVns{}, 
while the normalized cross sections are shown in units of 1/bin. 
The $\ptmax$ differential cross section 
measurements are performed by replacing the $\mll$ variable by the $\ptmax$ 
variable in the $\WW$ SR in the simultaneous fit. The measured absolute 
and normalized $\WZ$ differential cross sections in bins of $\mjj$ 
are shown in Fig.~\ref{fig:unf_wz_normalizedy}. The $\mjj$ differential cross section 
measurements are estimated by replacing the BDT variable by the $\mjj$ 
variable with 8 bins ([500, 650, 800, 1000, 1200, 1500, 1800, 2300, $\infty$]\GeV) in the $\WZ$ SR in the simultaneous fit. 
The measured cross section values agree with the theoretical predictions within the uncertainties.

\begin{table*}[htbp]
\centering
\topcaption{
The measured inclusive cross sections for the EW $\WW$, EW+QCD $\WW$, EW $\WZ$, EW+QCD $\WZ$, and QCD $\WZ$ 
processes and the theoretical predictions with \MGvATNLO at LO. The EW processes include the corresponding interference contributions. 
The theoretical uncertainties include statistical, PDF, and scale uncertainties. Predictions with applying the $\mathcal{O}(\alpS\alpha^6)$ 
and $\mathcal{O}(\alpha^7)$ corrections to the \MGvATNLO LO cross sections, as described in the text, are also shown. The predictions of the QCD 
$\WW$ and $\WZ$ processes do not include additional corrections. 
All reported values are in fb.\label{tab:default_fid_inc}}
\cmsTable{
\begin{tabular}{cccc}
\hline
\multirow{2}{*}{Process} & \multirow{2}{*}{$\sigma \, \mathcal{B}$ (fb)} & Theoretical prediction       & Theoretical prediction    \\
                         &                                               & without NLO corrections (fb) & with NLO corrections (fb) \\
\hline
\multirow{2}{*}{EW $\WW$}     &  $3.98 \pm 0.45$            &  \multirow{2}{*}{$3.93 \pm 0.57$} &  \multirow{2}{*}{$3.31 \pm 0.47$} \\
                              &  $0.37\stat \pm 0.25\syst$  &					&				    \\
\multirow{2}{*}{EW+QCD $\WW$} &  $4.42\pm 0.47$             &  \multirow{2}{*}{$4.34 \pm 0.69$} &  \multirow{2}{*}{$3.72 \pm 0.59$} \\
                              &  $0.39\stat \pm 0.25\syst$  &					&				    \\
\multirow{2}{*}{EW $\WZ$}     &  $1.81\pm 0.41$             &  \multirow{2}{*}{$1.41 \pm 0.21$} &  \multirow{2}{*}{$1.24 \pm 0.18$} \\
                              &  $0.39\stat \pm 0.14\syst$  &					&				    \\
\multirow{2}{*}{EW+QCD $\WZ$} &  $4.97\pm 0.46$             &  \multirow{2}{*}{$4.54 \pm 0.90$} &  \multirow{2}{*}{$4.36 \pm 0.88$} \\
                              &  $0.40\stat \pm 0.23\syst$  &					&				    \\
\multirow{2}{*}{QCD $\WZ$}    &  $3.15\pm 0.49$             &  \multirow{2}{*}{$3.12 \pm 0.70$} &  \multirow{2}{*}{$3.12 \pm 0.70$} \\
                              &  $0.45\stat \pm 0.18\syst$  &					&				    \\
\hline
\end{tabular}
}
\end{table*}

\subsection{Limits on anomalous quartic gauge couplings}

The events in the $\WW$ and $\WZ$ SRs are used to constrain aQGCs 
in the effective field theory (EFT) framework~\cite{Degrande:2012wf}. Nine independent 
charge-conjugate and parity conserving dimension-8 effective operators are 
considered~\cite{Eboli:2006wa}. The S0 and S1 operators are constructed from 
the covariant derivative of the Higgs doublet. The T0, T1, and T2 operators are 
constructed from the SU$_\mathrm{L}$(2) gauge fields. The mixed operators 
M0, M1, M6, and M7 involve the SU$_\mathrm{L}$(2) gauge fields and the Higgs doublet. 

A nonzero aQGC enhances the production cross section at large masses of the 
$\WW$ and $\WZ$ systems with respect to the SM prediction. 
The diboson transverse mass, defined as
\begin{equation}
\mT(\PV\PV) = \sqrt{{\biggl(\sum\nolimits_i E_{i}\biggr)^2-\biggl(\sum\nolimits_i p_{z,i}\biggr)^2}},
\label{eq:mtvv}
\end{equation}
where $E_{i}$ and $p_{z,i}$ are the energies and longitudinal components of the momenta of the leptons 
and neutrinos from the decay of the gauge bosons in the event, is used in the fit for 
both $\WW$ and $\WZ$ processes. The four-momentum of the neutrino system is defined using the $\ptvecmiss$, 
assuming that the values of the longitudinal component of the momentum and the invariant mass are zero.

\begin{figure}[htbp]
\centering
\includegraphics[width=0.49\textwidth]{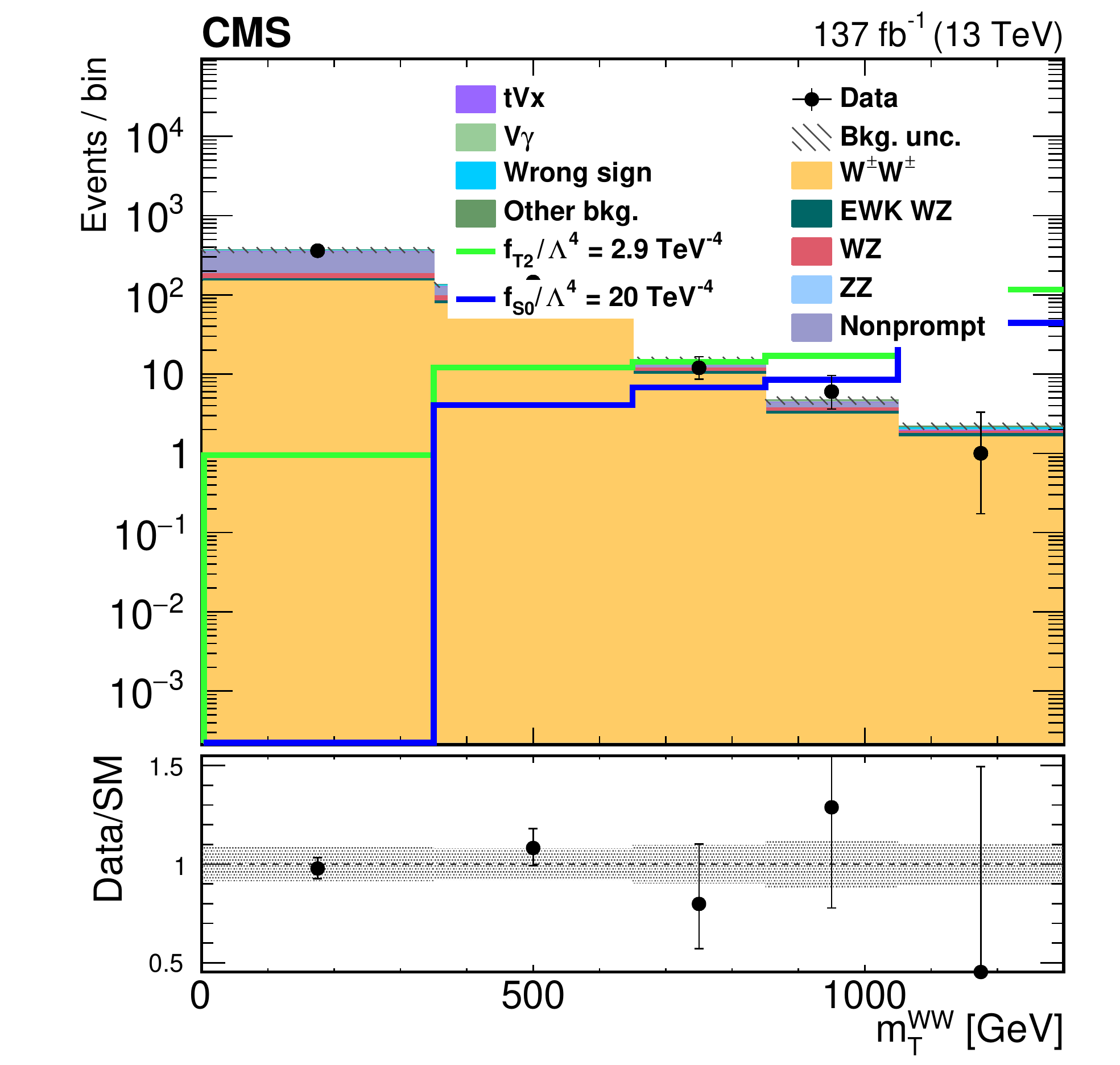}
\includegraphics[width=0.49\textwidth]{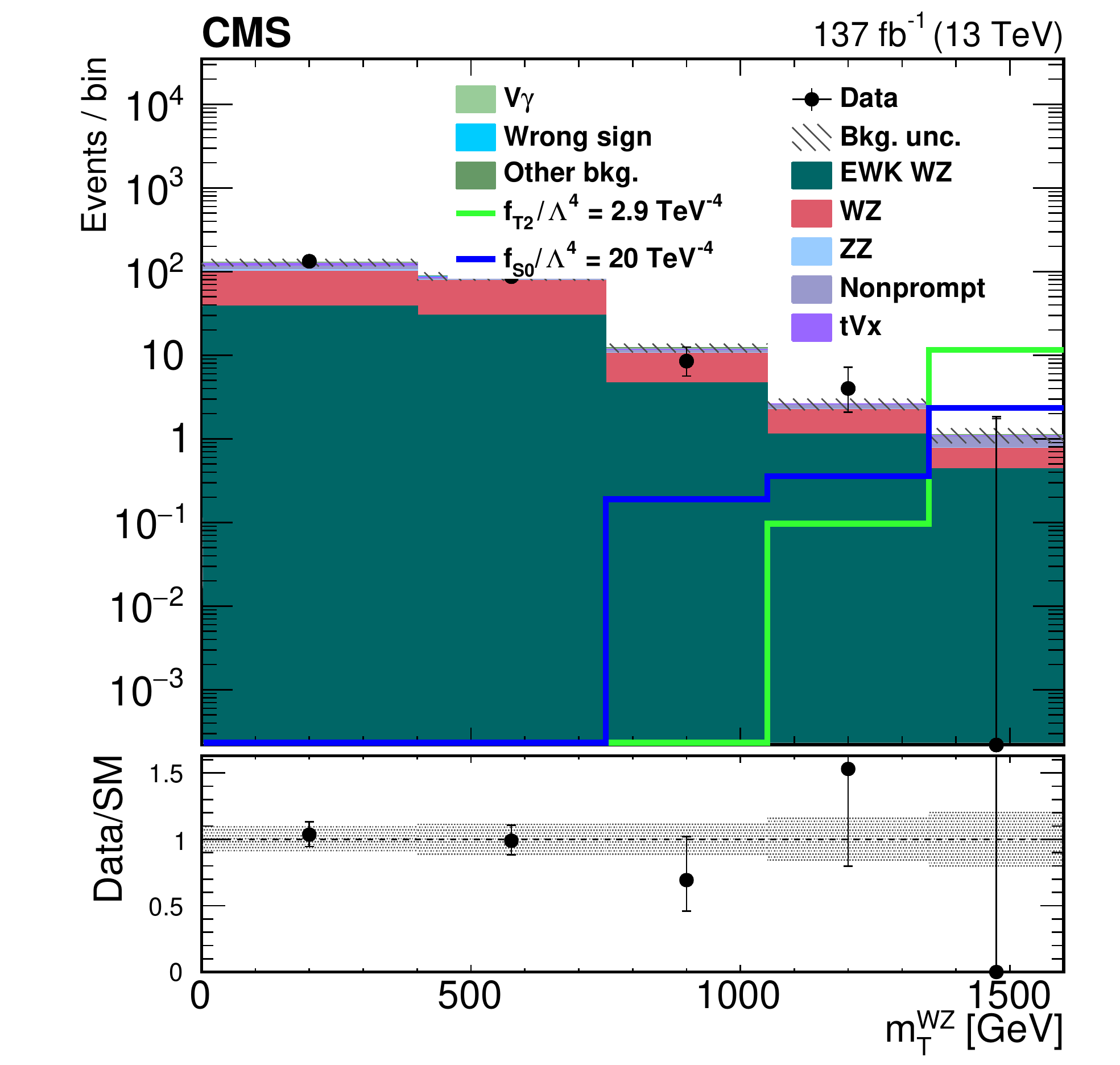}
\caption{Distributions of $\mT(\PW\PW)$ (\cmsLeft) in the $\WW$ SR and $\mT(\PW\cPZ)$ (\cmsRight) in the $\WZ$ SR. 
The gray bands include uncertainties from the predicted yields. The SM predicted yields are shown with their best fit 
normalizations from the corresponding fits. The contribution of the QCD $\WW$ process is included together with the EW $\WW$ process.
 The overflow is included in the last bin. The bottom panel 
in each figure shows the ratio of the number of events observed in data to the total SM prediction. 
The solid lines show the signal predictions for two illustrative aQGC parameters.\label{fig:ssww_aqgc}}
\end{figure}

A two-dimensional distribution is used in the fit for the $\WW$ process 
with 5 bins in $\mT(\PW\PW)$ ([0, 350, 650, 850, 1050, $\infty$]\GeV) and 4 bins in 
$\mjj$ ([500, 800, 1200, 1800, $\infty$]\GeV). The SM $\WZ$ 
contribution is considered to be background. Similarly, a two-dimensional distribution is 
used in the fit for the $\WZ$ process with 5 bins in $\mT(\PW\cPZ)$  
([0, 400, 750, 1050, 1350, $\infty$]\GeV) and 2 bins in $\mjj$ ([500, 1200, $\infty$]\GeV). 
The $\mjj$ distribution is used for the nonprompt lepton, $\tZq$, and $\cPZ\cPZ$ CRs in both fits 
with 4 bins ([500, 800, 1200, 1800, $\infty$]\GeV). The distributions of $\mT(\PV\PV)$ 
in the $\WW$ and $\WZ$ SRs are shown in Fig.~\ref{fig:ssww_aqgc}. 

No excess of events with respect to the SM background predictions is observed. 
The observed and expected $95\%$ confidence level (\CL) lower and upper limits 
on the aQGC parameters $f/\Lambda^4$, where $f$ is the dimensionless coefficient 
of the given operator and $\Lambda$ is the energy scale of new physics, are derived 
from a modified frequentist approach with the \CLs criterion~\cite{Junk,Read} 
and asymptotic results for the test statistic~\cite{CLs}. 
The expected cross section depends quadratically on aQGC, therefore the expected yields are 
calculated from a parabolic interpolation from the discrete coupling parameters of the 
simulated signals. Table~\ref{tab:VBS_aQGC} shows the individual lower and upper 
limits for the coefficients of the T0, T1, T2, M0, M1, M6, M7, S0, and S1 operators 
obtained by setting all other aQGCs parameters to zero for the $\WW$ 
and $\WZ$ channels, and their combination. The results are sensitive to the number of data events with large $\mT(\PV\PV)$ values.
These results are about a factor of two more restrictive than the previous analyses 
of the leptonic decay modes of the $\WW$ and $\WZ$ processes~\cite{Sirunyan:2017ret,Sirunyan:2019ksz}. However, the results are less restrictive than the analysis using the semileptonic final states~\cite{Sirunyan:2019der}. No unitarization procedure is applied to obtain these results.

\begin{table*}[htbp]
\centering
\topcaption{
Observed and expected lower and upper 95\% \CL limits on the parameters of the quartic 
operators T0, T1, T2, M0, M1, M6, M7, S0, and S1 in $\WW$ and $\WZ$ 
channels, obtained without using any unitarization procedure. The last two columns 
show the observed and expected limits for the combination of the 
$\WW$ and $\WZ$ channels. Results are obtained by setting all other aQGCs parameters to zero.\label{tab:VBS_aQGC}}
\cmsTable{
\begin{tabular}{ccccccc}
\hline
& Observed ($\WW$)  & Expected ($\WW$) & Observed ($\WZ$) & Expected ($\WZ$) & Observed & Expected  \\
& ($\TeVns^{-4}$)   & ($\TeVns^{-4}$)  & ($\TeVns^{-4}$)   & ($\TeVns^{-4}$) & ($\TeVns^{-4}$)   & ($\TeVns^{-4}$)  \\
\hline
$f_{\mathrm{T0}} / \Lambda^{4}$ & [-0.28, 0.31] & [-0.36, 0.39] & [-0.62, 0.65] & [-0.82, 0.85] & [-0.25, 0.28] & [-0.35, 0.37] \\
$f_{\mathrm{T1}} / \Lambda^{4}$ & [-0.12, 0.15] & [-0.16, 0.19] & [-0.37, 0.41] & [-0.49, 0.55] & [-0.12, 0.14] & [-0.16, 0.19] \\
$f_{\mathrm{T2}} / \Lambda^{4}$ & [-0.38, 0.50] & [-0.50, 0.63] & [-1.0 , 1.3]  & [-1.4,  1.7]  & [-0.35, 0.48] & [-0.49, 0.63] \\
$f_{\mathrm{M0}} / \Lambda^{4}$ & [-3.0, 3.2]   & [-3.7, 3.8]   & [-5.8, 5.8]   & [-7.6, 7.6]   & [-2.7, 2.9]   & [-3.6, 3.7]   \\
$f_{\mathrm{M1}} / \Lambda^{4}$ & [-4.7, 4.7]   & [-5.4, 5.8]   & [-8.2, 8.3]   & [-11, 11]     & [-4.1, 4.2]   & [-5.2, 5.5]   \\
$f_{\mathrm{M6}} / \Lambda^{4}$ & [-6.0, 6.5]   & [-7.5, 7.6]   & [-12, 12]     & [-15, 15]     & [-5.4, 5.8]   & [-7.2, 7.3]   \\
$f_{\mathrm{M7}} / \Lambda^{4}$ & [-6.7, 7.0]   & [-8.3, 8.1]   & [-10, 10]     & [-14, 14]     & [-5.7, 6.0]   & [-7.8, 7.6]   \\
$f_{\mathrm{S0}} / \Lambda^{4}$ & [-6.0, 6.4]   & [-6.0, 6.2]   & [-19, 19]     & [-24, 24]     & [-5.7, 6.1]   & [-5.9, 6.2]   \\
$f_{\mathrm{S1}} / \Lambda^{4}$ & [-18, 19]     & [-18, 19]     & [-30, 30]     & [-38, 39]     & [-16, 17]     & [-18, 18]     \\
\hline
\end{tabular}
}
\end{table*}

The EFT is not a complete model and the presence of 
nonzero aQGCs will violate tree-level unitarity at sufficiently high energy. 
More physical limits can be obtained by cutting the EFT integration at the 
unitarity limit and adding the expected SM contribution for generated events 
with $\PV\PV$ invariant masses above the unitarity limit~\cite{Kalinowski:2018oxd}. 
The unitarity limits for each aQGC parameter, typically about 1.5\TeV, are calculated using 
\textsc{vbfnlo} 1.4.0~\cite{Arnold:2008rz,Arnold:2011wj,Baglio:2014uba} 
after applying the appropriate Wilson coefficient conversion factors. 
Table~\ref{tab:VBS_aQGC_clip} shows the individual lower and upper limits for the 
coefficients of the T0, T1, T2, M0, M1, M6, M7, S0, and S1 operators by cutting off 
the EFT expansion at the unitarity limit. These limits are significantly less 
stringent compared with the limits in Table~\ref{tab:VBS_aQGC}, where the unitarity 
violation is not considered.  

\begin{table*}[htbp]
\centering
\topcaption{
Observed and expected lower and upper 95\% \CL limits on the parameters of the quartic operators T0, T1, T2, M0, M1,
M6, M7, S0, and S1 in $\WW$ and $\WZ$ channels by cutting the EFT expansion at the 
unitarity limit. The last two columns show the observed and expected limits for the combination of 
the $\WW$ and $\WZ$ channels. Results are obtained by setting all other aQGCs parameters to zero.\label{tab:VBS_aQGC_clip}}
\cmsTable{
\begin{tabular}{ccccccc}
\hline
& Observed ($\WW$)  & Expected ($\WW$) & Observed ($\WZ$) & Expected ($\WZ$) & Observed & Expected  \\
& ($\TeVns^{-4}$)   & ($\TeVns^{-4}$)  & ($\TeVns^{-4}$)   & ($\TeVns^{-4}$) & ($\TeVns^{-4}$)   & ($\TeVns^{-4}$)  \\
\hline
$f_{\mathrm{T0}} / \Lambda^{4}$	&	[-1.5, 2.3]	&	[-2.1, 2.7]	&	[-1.6, 1.9]	&	[-2.0, 2.2]	&	[-1.1, 1.6]	&	[-1.6, 2.0]	\\
$f_{\mathrm{T1}} / \Lambda^{4}$	&	[-0.81, 1.2]	&	[-0.98, 1.4]	&	[-1.3, 1.5]	&	[-1.6, 1.8]	&	[-0.69, 0.97]	&	[-0.94, 1.3]	\\
$f_{\mathrm{T2}} / \Lambda^{4}$	&	[-2.1,  4.4] 	&	[-2.7, 5.3]	&	[-2.7, 3.4]	&	[-4.4, 5.5]	&	[-1.6, 3.1]	&	[-2.3, 3.8]	\\
$f_{\mathrm{M0}} / \Lambda^{4}$	&	[-13, 16]	&	[-19, 18]	&	[-16, 16]	&	[-19, 19]	&	[-11, 12]	&	[-15, 15]	\\
$f_{\mathrm{M1}} / \Lambda^{4}$	&	[-20, 19]	&	[-22, 25]	&	[-19, 20]	&	[-23, 24]	&	[-15, 14]	&	[-18, 20]	\\
$f_{\mathrm{M6}} / \Lambda^{4}$	&	[-27, 32]	&	[-37, 37]	&	[-34, 33]	&	[-39, 39]	&	[-22, 25]	&	[-31, 30]	\\
$f_{\mathrm{M7}} / \Lambda^{4}$	&	[-22, 24]	&	[-27, 25]	&	[-22, 22]	&	[-28, 28]	&	[-16, 18]	&	[-22, 21]	\\
$f_{\mathrm{S0}} / \Lambda^{4}$	&	[-35, 36]	&	[-31, 31]	&	[-83, 85]	&	[-88, 91]	&	[-34, 35]	&	[-31, 31]	\\
$f_{\mathrm{S1}} / \Lambda^{4}$	&	[-100, 120]	&	[-100, 110]	&	[-110, 110]	&	[-120, 130]	&  	[-86, 99]	&	[-91, 97]	\\
\hline
\end{tabular}
}
\end{table*}

\section{Summary}
\label{sec:summary}

The production cross sections of $\WZ$ and same-sign $\PW\PW$ boson pairs in 
association with two jets are measured in proton-proton collisions at a 
center-of-mass energy of 13\TeV. The data sample corresponds 
to an integrated luminosity of 137\fbinv, collected 
with the CMS detector during 2016--18. The measurements are performed in the leptonic 
decay  modes $\PW^\pm\cPZ \to \ell^\pm\PGn\ell'^\pm\ell'^\mp$ and 
$\WW \to \ell^\pm\PGn\ell'^\pm\PGn$, where $\ell, \ell' = \Pe$, $\PGm$. 
An observation of electroweak production of $\WZ$ boson pairs is reported with 
an observed (expected) significance of 6.8 (5.3) standard deviations. Differential 
cross sections as functions of the invariant masses of the jet 
and charged lepton pairs, as well as the leading-lepton transverse momentum, are 
measured for $\WW$ production and are compared to the 
standard model predictions. Differential cross sections as a function of the 
invariant mass of the jet pair are also measured for $\WZ$ production. 
Stringent limits are set in the framework of effective field theory, with and without 
consideration of tree-level unitarity violation, on the dimension-8 
operators T0, T1, T2, M0, M1, M6, M7, S0, and S1.

\begin{acknowledgments}
  We congratulate our colleagues in the CERN accelerator departments for the excellent performance of the LHC and thank the technical and administrative staffs at CERN and at other CMS institutes for their contributions to the success of the CMS effort. In addition, we gratefully acknowledge the computing centers and personnel of the Worldwide LHC Computing Grid for delivering so effectively the computing infrastructure essential to our analyses. Finally, we acknowledge the enduring support for the construction and operation of the LHC and the CMS detector provided by the following funding agencies: BMBWF and FWF (Austria); FNRS and FWO (Belgium); CNPq, CAPES, FAPERJ, FAPERGS, and FAPESP (Brazil); MES (Bulgaria); CERN; CAS, MoST, and NSFC (China); COLCIENCIAS (Colombia); MSES and CSF (Croatia); RPF (Cyprus); SENESCYT (Ecuador); MoER, ERC IUT, PUT and ERDF (Estonia); Academy of Finland, MEC, and HIP (Finland); CEA and CNRS/IN2P3 (France); BMBF, DFG, and HGF (Germany); GSRT (Greece); NKFIA (Hungary); DAE and DST (India); IPM (Iran); SFI (Ireland); INFN (Italy); MSIP and NRF (Republic of Korea); MES (Latvia); LAS (Lithuania); MOE and UM (Malaysia); BUAP, CINVESTAV, CONACYT, LNS, SEP, and UASLP-FAI (Mexico); MOS (Montenegro); MBIE (New Zealand); PAEC (Pakistan); MSHE and NSC (Poland); FCT (Portugal); JINR (Dubna); MON, RosAtom, RAS, RFBR, and NRC KI (Russia); MESTD (Serbia); SEIDI, CPAN, PCTI, and FEDER (Spain); MOSTR (Sri Lanka); Swiss Funding Agencies (Switzerland); MST (Taipei); ThEPCenter, IPST, STAR, and NSTDA (Thailand); TUBITAK and TAEK (Turkey); NASU (Ukraine); STFC (United Kingdom); DOE and NSF (USA). 
 
  \hyphenation{Rachada-pisek} Individuals have received support from the Marie-Curie program and the European Research Council and Horizon 2020 Grant, contract Nos.\ 675440, 752730, and 765710 (European Union); the Leventis Foundation; the A.P.\ Sloan Foundation; the Alexander von Humboldt Foundation; the Belgian Federal Science Policy Office; the Fonds pour la Formation \`a la Recherche dans l'Industrie et dans l'Agriculture (FRIA-Belgium); the Agentschap voor Innovatie door Wetenschap en Technologie (IWT-Belgium); the F.R.S.-FNRS and FWO (Belgium) under the ``Excellence of Science -- EOS" -- be.h project n.\ 30820817; the Beijing Municipal Science \& Technology Commission, No. Z191100007219010; the Ministry of Education, Youth and Sports (MEYS) of the Czech Republic; the Deutsche Forschungsgemeinschaft (DFG) under Germany's Excellence Strategy -- EXC 2121 ``Quantum Universe" -- 390833306; the Lend\"ulet (``Momentum") Program and the J\'anos Bolyai Research Scholarship of the Hungarian Academy of Sciences, the New National Excellence Program \'UNKP, the NKFIA research grants 123842, 123959, 124845, 124850, 125105, 128713, 128786, and 129058 (Hungary); the Council of Science and Industrial Research, India; the HOMING PLUS program of the Foundation for Polish Science, cofinanced from European Union, Regional Development Fund, the Mobility Plus program of the Ministry of Science and Higher Education, the National Science Center (Poland), contracts Harmonia 2014/14/M/ST2/00428, Opus 2014/13/B/ST2/02543, 2014/15/B/ST2/03998, and 2015/19/B/ST2/02861, Sonata-bis 2012/07/E/ST2/01406; the National Priorities Research Program by Qatar National Research Fund; the Ministry of Science and Education, grant no. 14.W03.31.0026 (Russia); the Tomsk Polytechnic University Competitiveness Enhancement Program and ``Nauka" Project FSWW-2020-0008 (Russia); the Programa Estatal de Fomento de la Investigaci{\'o}n Cient{\'i}fica y T{\'e}cnica de Excelencia Mar\'{\i}a de Maeztu, grant MDM-2015-0509 and the Programa Severo Ochoa del Principado de Asturias; the Thalis and Aristeia programs cofinanced by EU-ESF and the Greek NSRF; the Rachadapisek Sompot Fund for Postdoctoral Fellowship, Chulalongkorn University and the Chulalongkorn Academic into Its 2nd Century Project Advancement Project (Thailand); the Kavli Foundation; the Nvidia Corporation; the SuperMicro Corporation; the Welch Foundation, contract C-1845; and the Weston Havens Foundation (USA). 
\end{acknowledgments}

\bibliography{auto_generated}
\cleardoublepage \appendix\section{The CMS Collaboration \label{app:collab}}\begin{sloppypar}\hyphenpenalty=5000\widowpenalty=500\clubpenalty=5000\vskip\cmsinstskip
\textbf{Yerevan Physics Institute, Yerevan, Armenia}\\*[0pt]
A.M.~Sirunyan$^{\textrm{\dag}}$, A.~Tumasyan
\vskip\cmsinstskip
\textbf{Institut f\"{u}r Hochenergiephysik, Wien, Austria}\\*[0pt]
W.~Adam, F.~Ambrogi, T.~Bergauer, M.~Dragicevic, J.~Er\"{o}, A.~Escalante~Del~Valle, R.~Fr\"{u}hwirth\cmsAuthorMark{1}, M.~Jeitler\cmsAuthorMark{1}, N.~Krammer, L.~Lechner, D.~Liko, T.~Madlener, I.~Mikulec, F.M.~Pitters, N.~Rad, J.~Schieck\cmsAuthorMark{1}, R.~Sch\"{o}fbeck, M.~Spanring, S.~Templ, W.~Waltenberger, C.-E.~Wulz\cmsAuthorMark{1}, M.~Zarucki
% ...
\vskip\cmsinstskip
\dag: Deceased\\
1:  Also at Vienna University of Technology, Vienna, Austria\\
% ...
\end{sloppypar}
%%% END EDITABLE REGION %%%
\end{document}